\documentclass[10pt,aps,prl,twocolumn,superscriptaddress,amsmath,amssymb,longbibliography,notitlepage]{revtex4-2}
\usepackage{graphicx,physics,mathtools}
\usepackage[usenames,dvipsnames,svgnames,table]{xcolor}
\usepackage[unicode=true, pdfusetitle, bookmarks=true, bookmarksnumbered=false, bookmarksopen=false, 
			pdfborder={0 0 0}, backref=false, colorlinks=true]{hyperref}
\hypersetup{linkcolor=NavyBlue,urlcolor=NavyBlue,citecolor=NavyBlue}

\newcommand{\est}{\hat}		 	
\newcommand{\qfi}{\mathcal{F}}  
\newcommand{\povm}{M}			
\newcommand{\hs}{\mathcal{H}}	
\newcommand{\idop}{\openone}    

\begin{document}

\title{Incorporating Heisenberg's Uncertainty Principle into Quantum Multiparameter Estimation}
\author{Xiao-Ming Lu}
\email{lxm@hdu.edu.cn}
\homepage{http://xmlu.me}
\affiliation{Department of Physics, Hangzhou Dianzi University, Hangzhou 310018, China}

\author{Xiaoguang Wang}
\email{xgwang1208@zju.edu.cn}
\affiliation{Zhejiang Institute of Modern Physics and Department of Physics, Zhejiang University, Hangzhou 310027, China}

\begin{abstract}
The quantum multiparameter estimation is very different from the classical multiparameter estimation due to Heisenberg's uncertainty principle in quantum mechanics.
When the optimal measurements for different parameters are incompatible, they cannot be jointly performed.
We find a correspondence relationship between the inaccuracy of a measurement for estimating the unknown parameter with the measurement error in the context of measurement uncertainty relations. 
Taking this correspondence relationship as a bridge, we incorporate Heisenberg's uncertainty principle into quantum multiparameter estimation by giving a tradeoff relation between the measurement inaccuracies for estimating different parameters.
For pure quantum states, this tradeoff relation is tight, so it can reveal the true quantum limits on individual estimation errors in such cases.
We apply our approach to derive the tradeoff between attainable errors of estimating the real and imaginary parts of a complex signal encoded in coherent states and obtain the joint measurements attaining the tradeoff relation.
We also show that our approach can be readily used to derive the tradeoff between the errors of jointly estimating the phase shift and phase diffusion without explicitly parameterizing quantum measurements. 
\end{abstract}
\maketitle

The random nature of quantum measurement imposes ultimate limits on the precision of estimating unknown parameters with quantum systems.
Quantum parameter estimation theory has been developing for more than half a century to reveal and pursue the quantum-limited measurement~\cite{Helstrom1967,Helstrom1968,Yuen1973,Belavkin1976,Helstrom1976,Holevo1982,Personick1971,HayashiBook,Tsang2020a}. 
In classical parameter estimation theory, the Cram\'er-Rao bound (CRB) together with the asymptotic normality of the maximum likelihood estimator give a satisfactory approach to derive the asymptotically attainable accuracy of estimation, where the Fisher information matrix (FIM) plays a pivotal role~\cite{Fisher1922,Fisher1925,Cramer1946,Rao1945,Kay1993,Wasserman2010,Casella2002,Lehmann1998}.
The CRB and the FIM have been extended to quantum regime~\cite{Helstrom1967,Helstrom1968,Yuen1973,Belavkin1976,Helstrom1976,Holevo1982}, where not only estimators---data processing---but also quantum measurements are taken into consideration in optimization. 

For single parameter estimation, Helstrom's version of quantum CRB can be attained at large samples due to the asymptotic efficiency of adaptive measurements~\cite{Braunstein1994,Braunstein1996,Fujiwara2006,HayashiBook}.
However,  unlike the classical parameter estimation, the quantum CRB does not possess the asymptotic attainability in general for multiparameter estimation.
This can be understood as a consequence of the fact that the optimal measurements for different parameters may be incompatible in quantum mechanics so that they cannot be jointly performed according to Heisenberg's uncertainty principle (HUP)~\cite{Heisenberg1927,Busch2007}.
Many application scenarios, e.g., superresolution imaging~\cite{Tsang2016b,Tsang2019a}, quantum enhanced estimation of a magnetic field~\cite{Baumgratz2016,Hou2020}, and joint estimation of phase shift and phase diffusion~\cite{Vidrighin2014}, essentially belong to quantum multiparameter estimation problems. 
Therefore, the characterization of the quantum-limited bound on the estimation errors is of great importance to many practical applications of quantum estimation.
Nevertheless, it is still challenging to derive, characterize, and understand the quantum limit on accuracies of the multiparameter estimation~\cite{Tsang2020a,Carollo2019a,Rubio2018,Albarelli2019,Tsang2019f,Albarelli2019a,Sidhu2020,Lu2020a,Sidhu,Ragy2016,Li2016,Zhu2018,Suzukia2016,Suzuki2019,Suzuki2019a,Kull2020,Carollo2020,Li2017,Liu2020,Gill2000,Nagaoka2005a,Matsumoto2002}.

Due to the difficulty in identifying the boundary between the forbidden and permissible regions of error composition, as a compromise, many prior error bounds are formulated in terms of the weighted mean errors of estimation~\cite{Helstrom1967,Helstrom1968,Yuen1973,Holevo1982,Tsang2020a,Carollo2019a,Rubio2018,Albarelli2019,Tsang2019f,Albarelli2019a,Sidhu2020,Lu2020a,Sidhu,Ragy2016,Li2016,Zhu2018,Suzukia2016,Suzuki2019,Suzuki2019a,Kull2020,Carollo2020,Li2017,Liu2020,Gill2000,Nagaoka2005a,Matsumoto2002}. 
The most powerful lower bound on weighted mean errors up to now is the Holevo bound~\cite{Holevo1982,Albarelli2019}, which is asymptotically attainable by collective measurements on a large number of identical samples~\cite{Gill2013,Hayashi2008,Kahn2009,Yamagata2013,Guta2006}.
However, the Holevo bound itself contains an optimization over a set of special operators so that is difficult to be calculated~\cite{Albarelli2019}. 
Remarkably, Carollo {\it et al}. derived an upper bound on the discrepancy ratio between the Holevo bound and Helstrom's version of quantum CRB through a quantity measuring the incompatibility regarding different parameters~\cite{Carollo2019a}.
With these lower and upper bounds, we can reveal the quantum limits on weighted mean errors, nevertheless, are still difficult to completely identify the tradeoff curve/surface regarding the attainable errors for estimating different parameters~\cite{Kull2020,Lu2020a}.
It is still unclear how the HUP affects the boundary of the attainable errors.

In this work, we tackle the problem of completely identifying the boundary of the attainable errors of estimating multiple parameters by \emph{directly} incorporating the HUP into quantum multiparameter estimation.
We define the regret of Fisher information for a quantum measurement that is used to estimate an unknown parameter and shall derive the following correspondence relation:
\[
	\mathsf{Information\ Regret \leftrightarrow Measurement\ Error}.
\]
Taking this relationship as a bridge, we obtain tradeoff relations between the information regrets for different parameters through Branciard's and Ozawa's versions of measurement uncertainty relations in terms of the state-dependent measurement error defined by Ozawa~\cite{Ozawa2003,Ozawa2004a,Hall2004,Weston2013,Branciard2013,Lu2014}.
This tradeoff relation is tight for pure quantum states, so it can faithfully reveal the quantum limits on multiparameter estimation errors with pure quantum states. 
We shall apply the regret tradeoff relation to the coherent state estimation and the joint estimation of phase shift and phase diffusion.

Let us start with a brief introduction on quantum multiparameter estimation.
Let \( \theta = (\theta_1,\theta_2,\ldots,\theta_n) \in \mathbb{R}^n \) be an unknown vector parameter, which can be estimated via observing  a quantum system.
The state of the quantum system depends on the true value of \(\theta\) and is described by a parametric density operator \(\rho_\theta\).
The quantum measurement can be characterized by a positive-operator-valued measure (POVM) \( \povm = \{\povm_x | \povm_x\geq 0, \sum_x \povm_x=\openone \} \), where \(x\) denotes the outcome and \( \openone \) is the identity operator.
Denote the estimator for \(\theta\) by \( \est \theta = (\est\theta_1, \est\theta_2, \ldots, \est\theta_n) \), which is a map from the observation data to the estimates.
The estimation error can be characterized by the error-covariance matrix defined by its entries
\(
 	\mathcal{E}_{jk} = \mathbb E_\theta [(\est\theta_j - \theta_j) (\est\theta_k - \theta_k)],
\)
where the expectation \( \mathbb E_\theta[\bullet] \) is taken with respect to the observation data with the joint probability mass function \( p_\theta(x_1,x_2,\ldots,x_\nu) = \prod_{j=1}^\nu\tr(\povm_{x_\nu} \rho_\theta) \) with \(\nu\) being the number of experimental runs with independent and identically distributed samples.
The error-covariance matrix of any unbiased estimator \( \est\theta \) obeys the CRB \(\mathcal{E} \geq \nu^{-1} F^{-1}\) in the sense that the matrix \( \mathcal{E} - \nu^{-1} F^{-1} \) is positive semi-definite~\cite{Kay1993,Lehmann1998,Casella2002,Wasserman2010}, where \(F\) is the (classical) FIM for a single experimental run and defined by
\begin{equation}
	F_{jk} = \mathbb E_\theta\qty[
        \frac{\partial\ln p_\theta(x)}{\partial \theta_j} \frac{\partial\ln p_\theta(x)}{\partial \theta_k} 
	]
\end{equation} 
with \(p_\theta(x)=\tr(M_x\rho_\theta)\).
The CRB is asymptotically attainable by the maximum likelihood estimator~\cite{Fisher1922,Fisher1925}, whose distribution at a large \(\nu\) is approximate to a multivariate normal distribution with the mean being the true value of \(\theta\) and the covariance matrix being \( \nu^{-1} F^{-1} \), according to the central limit theorem~\cite[Theorem 9.27]{Wasserman2010}.

The FIM depends on the quantum measurement via \(p_\theta(x) = \tr(\povm_x\rho_\theta)\), so does the CRB.
We use \(F(\povm)\) to explicitly indicate the dependence of \(F\) on a POVM \(M\).
Quantum parameter estimation takes into consideration the optimization over quantum measurements. 
For any quantum measurement, the FIM is bounded by the following matrix inequality:~\cite{Braunstein1994,Hiai2014}
\begin{equation}\label{eq:bc}
	F(\povm) \leq \qfi, 
\end{equation}
where \(\qfi\) is the so-called quantum FIM, also known as the Helstrom information matrix~\cite{Helstrom1967,Helstrom1968}.
The quantum FIM is the real part of a Hermitian matrix \(\mathcal Q\) (i.e., \(\qfi = \Re\mathcal Q\)) defined by
\begin{equation}\label{eq:Q}
    \mathcal{Q}_{jk} = \tr ( L_j L_k \rho_\theta ),
\end{equation}
where \(L_j\), the symmetric logarithmic derivative (SLD) operator for \(\theta_j\), is a Hermitian operator satisfying \((L_j\rho_\theta + \rho_\theta L_j) / 2= \pdv*{\rho_\theta}{\theta_j}\).
Combining Eq.~\eqref{eq:bc} with the CRB yields the quantum CRB \(\mathcal{E} \geq \nu^{-1} \mathcal{F}^{-1}\) for any quantum measurement and any unbiased estimator.
This quantum CRB was first obtained by Helstrom with a different method~\cite{Helstrom1967,Helstrom1968}.

To characterize the inaccuracy of a quantum measurement for multiparameter estimation, we here define the regret of Fisher information by
\begin{align}
	R(\povm) &= \qfi - F(\povm).
\end{align}
This matrix \(R(\povm)\) is positive semi-definite due to Eq.~\eqref{eq:bc} and real symmetric as both the quantum and classical FIMs are real symmetric according to their definitions.
For single-parameter estimation, Braunstein and Caves proved that the classical Fisher information can equal the quantum Fisher information with an optimal quantum measurement~\cite{Braunstein1994} and thus the regret \(R(\povm)\) thereof vanishes.
In the multiparameter setting, for any column vector \( v\in\mathbb R^n \), there exist a quantum measurement \(\povm\) such that \(v^\top R(\povm) v = 0\), where \(\top\) denotes matrix transpose.
This is because \( v^\top F(\povm) v \) and \( v^\top \qfi v \) can be interpreted as the classical and quantum Fisher information, respectively, about a parameter \( \varphi \) satisfying \(\pdv*{\varphi}=\sum_j v_j \pdv*{\theta_j}\).
The POVM \(M\) making \(v R(M) v^\top\) vanish can be considered as an optimal measurement for estimating \(\varphi\) and in general depends on \(v\). 
For different parameters, the optimal measurement may be different and even incompatible.
Consequently, the entries of \(R(\povm)\) in general cannot simultaneously vanish, which is a manifestation of HUP.
In what follows, we shall give a quantitative characterization of the mechanism in which the HUP affects the regret matrix of Fisher information.


Define by \(\Delta_j = \sqrt{R_{jj} / \qfi_{jj}} \) the normalized-square-root regret of Fisher information with respect to \(\theta_j\).
Note that \(\Delta_j\) takes value in the interval \([0,1]\).
Our main result is the following tradeoff relation:
\begin{equation}\label{eq:main}
    \Delta_j^2 + \Delta_k^2 + 2 \sqrt{1-c_{jk}^2} \Delta_j \Delta_k \geq c_{jk}^2,
\end{equation}
where \(c_{jk}\) is a real number given by
\begin{equation}
	c_{jk} = \frac{ |\Im\mathcal Q_{jk}| }{ \sqrt{\Re\mathcal Q_{jj} \Re\mathcal Q_{kk}} }
	= \frac{|\Im\mathcal Q_{jk}|}{ \sqrt{\qfi_{jj} \qfi_{kk}} }
\end{equation}
with \(\mathcal Q_{jk}\) being given by Eq.~\eqref{eq:Q}.
For nonzero \(c_{jk}\), Eq.~\eqref{eq:main} describes the tradeoff between the regrets of Fisher information with respect to different parameters.
For a family \(\rho_\theta\) of pure states, the inequality Eq.~\eqref{eq:main} is tight, in the sense that there exists a quantum measurement \( M \) such that the equality in Eq.~\eqref{eq:main} holds;
In such a case, our result fully reflects the tradeoff between different regrets of Fisher information.
For mixed states \(\rho_\theta\), the inequality Eq.~\eqref{eq:main} can be tightened by replacing \(c_{jk}\) thereof by its variant
\begin{equation}
	\widetilde c_{jk}=\frac{
		\tr\abs{\sqrt{\rho_\theta} [L_j,L_k] \sqrt{\rho_\theta}}
	}{ 2 \sqrt{ \qfi_{jj} \qfi_{kk}} },
\end{equation}
where \(|X| = \sqrt{X^\dagger X}\) for an operator \(X\).
Note that the coefficient \(\widetilde c_{jk}\) is not less than \(c_{jk}\) for all quantum states and equal to \(c_{jk}\) for all pure states.
We also give the second form of the tradeoff relation in terms of the estimation errors:
\begin{align}\label{eq:main2}
	\gamma_j + \gamma_k
	- 2 \sqrt{1 - \widetilde c_{jk}^2} \sqrt{(1 - \gamma_j)(1 - \gamma_k)}
	\leq 2 - \widetilde c_{jk}^2,
\end{align}	
where we have defined \(\gamma_j = 1 / (\nu \mathcal{E}_{jj} \qfi_{jj})\) for simplicity.
The above inequality is a result of combining Eq.~\eqref{eq:main} with the classical CRB \(\mathcal E_{jj} \geq \nu^{-1} (F^{-1})_{jj} \geq 1 / (\nu F_{jj})\).

\begin{figure}[tb]
	\centering
	\includegraphics[]{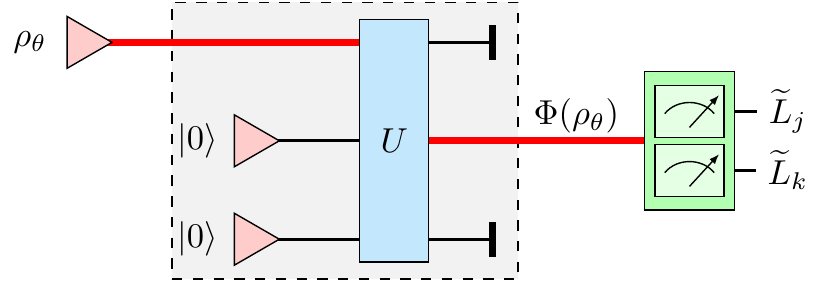}
	\caption{
		Unitary implementation (the dashed box) of the measurement channel.
		The thick red lines stand for the input and output ports.
		The commuting observables \(\widetilde L_j\) and \(\widetilde L_k\) can be jointly measured in the output state \(\Phi(\rho_\theta)\), which in the Heisenberg picture is equivalent to the joint measurement of a pair of commuting observables \(\mathcal L_j = U^\dagger (\idop_\mathrm{s} \otimes \widetilde L_j \otimes \idop_\mathrm{r}) U\) and \(\mathcal L_k = U^\dagger (\idop_\mathrm{s} \otimes \widetilde L_k \otimes \idop_\mathrm{r}) U\) in the initial state \(\rho_\mathrm{total} = \rho_\theta \otimes \dyad0 \otimes \dyad0\) of the entirety.
	}
	\label{fig:channel}
\end{figure}

We here outline the proof of Eq.~\eqref{eq:main} and leave the details in the Supplemental Material~\cite{SupplementalMaterial}.
Denote by \( \hs_\mathrm{s} \) the Hilbert space associated with the underlying quantum system.
For a given POVM \(\povm\) on \(\hs_\mathrm{s}\), we define a measurement channel \( \Phi(\rho) = \sum_x \tr(\povm_x \rho) \dyad x \), where \( \{\ket x\} \) is an orthonormal basis associated with the measurement outcomes \(x\)'s and span another Hilbert space \(\hs_\mathrm{r}\).
Note that the density operators \(\Phi(\rho_\theta)\) are always diagonal with the basis \( \{\ket x\} \).
As a result, the SLD operators of \(\Phi(\rho_\theta)\) are also diagonal with the basis \(\{\ket x\}\) and can be represented as
\begin{equation} \label{eq:diagonal_sld}
	\widetilde L_j  = \sum_x \pdv{\ln\tr(\povm_x \rho_\theta)}{\theta_j} \dyad x.
\end{equation}
The measurement channel \(\Phi\) can be implemented by a unitary operation \(U\) acting on \(\hs_\mathrm{s} \otimes \hs_\mathrm{r} \otimes \hs_\mathrm{r}\) such that 
\begin{equation}
	\Phi(\rho) = \tr_{1,3}\qty[U \qty(\rho \otimes \dyad0 \otimes \dyad0) U^\dagger]
\end{equation}
for all density operators \(\rho\) on \(\hs_\mathrm{s}\), where \(\tr_{1,3}\) denotes the partial trace over the first and third tensor factors of the Hilbert space and \(\ket0\) can be an arbitrary initial state~\cite[Chapter 2]{WolfBook}
(see Fig.~\ref{fig:channel} for a schematic illustration).
Using the techniques developed in Ref.~\cite{Lu2015}, we show that~\cite{SupplementalMaterial}
\begin{equation}\label{eq:bridge}
	R_{jj} = \tr[(\mathcal L_j - L_j\otimes \idop_\mathrm{r}\otimes\idop_\mathrm{r})^2 \rho_\mathrm{total}],
\end{equation}
where 
\(
	\mathcal L_j = U^\dagger (\idop_\mathrm{s} \otimes \widetilde L_j \otimes \idop_\mathrm{r}) U 
\)
with \(\idop_\mathrm{s}\) and \(\idop_\mathrm{r}\) being the identity operators on \(\hs_\mathrm{s}\) and \(\hs_\mathrm{r}\), respectively, and \(\rho_\mathrm{total} = \rho_\theta \otimes \dyad0 \otimes \dyad0\).

We observe that \(R_{jj}\) expressed in Eq.~\eqref{eq:bridge} is of the same form as the square of Ozawa's definition of measurement error~\cite{Ozawa2003,Ozawa2004a,Ozawa2019}, when taking \(L_j\) as the ideal observable we intend to measure and \(\mathcal L_j\) as the observable actually measured.
We list in Table~\ref{tab:correspondence} the correspondence relation between the parameter estimation scenario and the measurement error scenario.
Notice that the Hermitian operators \(\mathcal L_j\) and \(\mathcal L_k\) always commutes, as both \(\widetilde L_j\) and \(\widetilde L_k\) are diagonal with the basis \(\{\ket x\}\).
Therefore, the observables \(\mathcal L_j\) and \(\mathcal L_k\) can be jointly measured in quantum mechanics.
When two ideal observables \(L_j\) and \(L_k\) do not commute, it may be impossible to make their measurement errors, which equals the regrets \(R_{jj}\) and \(R_{kk}\) in our context, simultaneously vanish.
By invoking the measurement uncertainty relations~\cite{Ozawa2003,Ozawa2004a,Hall2004,Branciard2013,Lu2014,Ozaw2014} in terms of Ozawa's definition of measurement error, we can derive the tradeoff relation between the regrets of Fisher information with respect to different parameters.
Concretely, the inequality Eq.~\eqref{eq:main} follows from Branciard's version of measurement uncertainty relation, which is tight for pure states~\cite{Branciard2013}.
Using Ozawa's work on strengthening Branciard's inequality for mixed states~\cite{Ozaw2014}, the inequality Eq.~\eqref{eq:main} can be tightened through replacing \(c_{jk}\) by \(\widetilde c_{jk}\).

\begin{table}
	\caption{Correspondence relation.\label{tab:correspondence}}
	\begin{ruledtabular}
		\begin{tabular}{ll}
			Estimation-regret scenario & Measurement-error scenario \\
			\hline
			regret of Fisher information & measurement error \\
			SLD \(L_j\) of \(\rho_\theta\) & ideal observable \(L_j\) on \(\rho_\theta\) \\
			SLD \(\widetilde L_j\) of \(\Phi(\rho_\theta)\) & approximate observable on \(\Phi(\rho_\theta)\)\\
			\(\mathcal L_j=U^\dagger (\idop_\mathrm{s} \otimes \widetilde L_j \otimes \idop_\mathrm{r}) U\) & approximate observable on \(\rho_\theta\)
		\end{tabular}
	\end{ruledtabular}
\end{table}

It is worthy to point out that we do not designate the SLD operator as the ideal observable in reality to optimally estimate an individual parameter.
Although the eigenstates of the SLD operator, which possibly depend on the true value of the parameter, in principle constitute a measurement basis extracting the maximum Fisher information at a parameter point~\cite{Braunstein1994}, it is possible for some models to find a global optimal measurement that is independent of the parameter~\cite{Holevo1982,Braunstein1996}; 
A global optimal measurement is often more ideal than a local one for estimating the unknown parameter.

We can give an operational significance to the coefficients \(\widetilde c_{jk}\)  through the tradeoff relation Eq.~\eqref{eq:main} as follows.
If the QFI about a parameter \( \theta_j \) is exhaustively extracted by a quantum measurement \( M\), i.e., \(\Delta_j=0\), then it follows from Eq.~\eqref{eq:main} that the regret for any other parameter \(\theta_k\) obeys \(\Delta_k \geq \widetilde c_{jk}\).
That is, \(\widetilde c_{jk}\) is the lower bound on the residual regret for \(\theta_k\) when there is no regret for \(\theta_j\).
For pure states, this lower bounds \(c_{jk}\) can be attained as Eq.~\eqref{eq:main} is tight in such cases.
s

\begin{figure}[tb]
	\centering
	\includegraphics[]{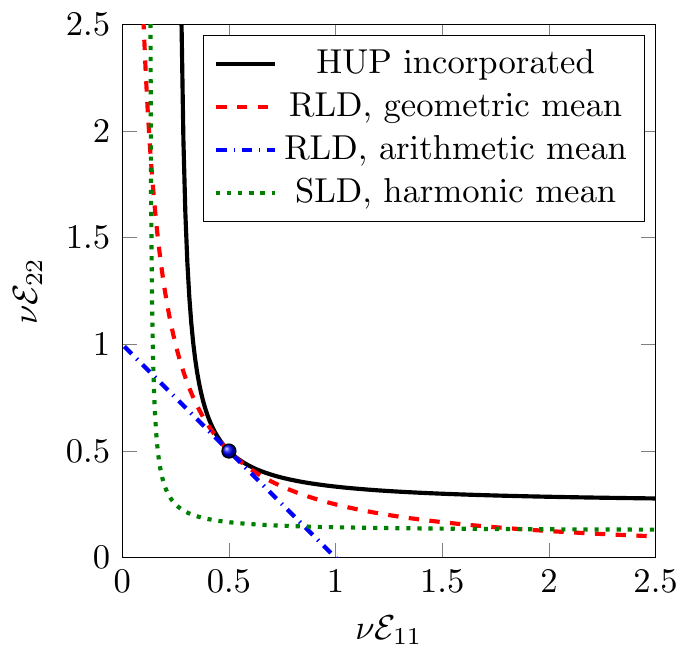}
	\caption{
		Mean-square errors of estimating the real and imaginary parts of a complex number \(\alpha\) encoded in a coherent state \(\ket\alpha\).
		The regions below the curves are forbidden by the corresponding inequalities.
		The black solid curve stands for the inequality Eq.~\eqref{eq:coherent_example} from the regret tradeoff relation.
		The other three curves are corresponding to the generalized-mean CRBs based on the SLD and the right logarithmic derivative (RLD) obtained in Ref.~\cite{Lu2020a}.
		Specifically, the red dashed curse stands for the RLD-based geometric-mean quantum CRB given by \(\sqrt{\mathcal E_{11} \mathcal E_{22}} \geq 1/(2\nu)\), the blue dash-dotted one stands for the RLD-based arithmetic-mean quantum CRB given by \((\mathcal E_{11} + \mathcal E_{22})/2 \geq 1/(2\nu)\), and the green dotted one stands for the SLD-based harmonic-mean quantum CRB given by \(2/(\mathcal E_{11}^{-1} + \mathcal E_{22}^{-1} ) \geq 1/(4\nu)\).
	}
	\label{fig:coherent_state_example}
\end{figure}

Let us now consider as an example the estimation of a complex number \(\alpha\) encoded in a coherent state~\cite{Glauber1963a} \(\ket\alpha\). 
The parameters of interest are the real and imaginary parts of \(\alpha\), i.e., \(\theta_1=\Re\alpha\) and \(\theta_2=\Im\alpha\).
After some algebras, we get \(\mathcal Q = 4 \smqty(1 & i\\ -i & 1)\) and thus \(c_{12} = 1\).
The regret tradeoff Eq.~\eqref{eq:main} then becomes \(\Delta_1^2 + \Delta_2^2 \geq 1\), which is equivalent to \(F_{11} + F_{22} \leq 4\) in terms of Fisher information or 
\begin{equation}\label{eq:coherent_example}
	\frac1{\nu\mathcal E_{11}} + \frac1{\nu\mathcal E_{22}} \leq 4
\end{equation}
in terms of estimation errors.
As shown in Fig.~\ref{fig:coherent_state_example}, Eq.~\eqref{eq:coherent_example} gives the most informative lower bound on the estimation error, compared with the error bounds that was previously investigated~\cite{Yuen1973,Holevo1982,Helstrom1976,Lu2020a}.

For this example, there exists a family of optimal (single-copy) measurements extracting the Fisher information such that the regret tradeoff relation in the above example, 
\(\Delta_1^2 + \Delta_2^2 \geq 1\), is saturated.
As a result, the error bound in Eq.~\eqref{eq:coherent_example} can be asymptotically attained.
We shall construct the optimal measurement as follows.
Denote by \(a\) the annihilation operators for the mode for which the coherent state is defined.
The measurements of the quadrature components \(Q = (a + a^\dagger) / 2\) and \(P = (a - a^\dagger) / (2i)\) are natural for estimating the coherent signal, as \(\ev{Q}{\alpha} = \Re\alpha\) and \(\ev{P}{\alpha}=\Im\alpha\).
Indeed, the maximum Fisher information about \(\theta_1\) and \(\theta_2\) can be obtained by measuring \(Q\) and \(P\), respectively, corresponding to either \(F_{11} = 4\) or \(F_{22} = 4\).
However, \(Q\) and \(P\) are not commuting so that they cannot be jointly measured.
It is known that we can jointly measure the commuting operators \(Q - Q'\) and \(P + P'\), where \(Q'\) and \(P'\) are the quadrature components of an ancillary mode (whose annihilation operator is denoted by \(a'\)) in the vacuum state, to  estimate the real and imaginary parts of \(\alpha\), see Refs.~\cite{Helstrom1976,Holevo1982,Yuen1973}.
This measurement strategy attains the minimum unweighted arithmetic mean error of estimation with \(F_{11} = F_{22} = 2\), see the blue circle in Fig.~\ref{fig:coherent_state_example}.
We show in the Supplemental Material~\cite{SupplementalMaterial} that other error combinations on the bound Eq.~\eqref{eq:coherent_example} can be asymptotically attained if we prepare the ancillary mode in a squeezed vacuum state \(\exp[\frac12(r a'^2 - r a'^{\dagger 2})] \ket0\) with \(r\in\mathbb R\). 
In such case, the extracted FIM can be tuned by changing \(r\) as \(F_{11} = 4 / (e^{2r} + 1)\), \(F_{22} = 4 e^{2r} / (e^{2r} + 1)\), and \(F_{12}=F_{21}=0\).
Moreover, the joint probability density function of outcomes of \(Q-Q'\) and \(P+P'\) are both Gaussian, so taking their sample means as the estimates for \(\theta_1\) and \(\theta_2\) asymptotically attains the classical CRB.

In our second example, we consider the joint estimation of phase shift and phase diffusion~\cite{Vidrighin2014}.
For a two-mode probe state, the parametric density operator can be effectively simplified as
\(\rho = \frac12(\op0 + \op1 + e^{-i\theta_1-\theta_2^2} \dyad{0}{1} + e^{i\theta_1-\theta_2^2} \dyad{1}{0} )\), 
where \(\theta_1\) stands for the phase shift and \(\theta_2\) the phase diffusion.
In Ref.~~\cite{Vidrighin2014}, Vidrighin \textit{et al.} obtained the tradeoff relation \(F_{11} / \qfi_{11} + F_{22} / \qfi_{22} \leq 1\) by explicitly parameterizing the rank-1 POVMs and then taking optimization.
We here show that Vidrighin \textit{et al.}'s tradeoff relations follows from our regret tradeoff relation Eq.~\eqref{eq:main} in a very easy way.
We only need to show \(\widetilde c_{12} = 1\) by a straightforward calculation according to its definition (see the Supplemental Material~\cite{SupplementalMaterial} for the details).
As a result, we get \(\Delta_1^2 + \Delta_2^2 \geq 1\), which is equivalent to Vidrighin \textit{et al.}'s tradeoff relation by recognizing \( \Delta_j^2 = 1 - F_{jj} / \qfi_{jj} \).

In conclusion, we have directly incorporated the HUP into quantum multiparameter estimation by deriving a tradeoff relation between the regrets of Fisher information about different parameters.
Unlike the quantum CRBs on scalar mean errors, the regrets tradeoff quantitatively characterizes how the HUP affects the combinations of estimation errors for multiple parameters.
The correspondence relationship we found between information regret and measurement error also, as a bonus, supplies an operational meaning to Ozawa's definition of the state-dependent measurement error, on which there exists a controversy for a long time~\cite{Busch2014c,Appleby2016,Ozawa2019}.

Our approach also opens a new perspective on quantum geometry.
The matrix \(\mathcal Q\) defined by Eq.~\eqref{eq:Q} is known as the quantum geometric tensor on the manifold of physical quantum state, up to an insignificant constant factor~\cite{Provost1980,Berry1989}.
The real part of \(\mathcal Q\)---the quantum FIM---gives a Riemannian metric on the manifolds of quantum states.
The imaginary part of \(\mathcal Q\) gives a curvature form of Berry's connection~\cite{Berry1989}, which has relations to the quantum FIM~\cite{Guo2016,Carollo2019a} and the density of quantum states~\cite{Xing2020}.
It is known that a zero curvature is necessary for the simultaneous vanishing of the regrets of Fisher information about different parameter~\cite{Matsumoto2002,Ragy2016,HayashiBook,Sidhu2020}.
Note that in our tradeoff relation, \(c_{jk}\) is the curvature divided by a scalar related to the entries of the quantum FIM. 
So our tradeoff relation quantitatively characterize the intricate mechanism in which the simultaneous reduction of the regrets of Fisher information about different parameters is restricted by a nonzero quantum curvature, which is indicated as 
\[
	\mathsf{Information\ Regret \leftarrow Quantum\ Curvature}. 
\]
Carollo \textit{et al.} has proposed an incompatibility index, which is similar to \(c_{jk}\), based on the ratio between the curvature and the quantum FIM as a figure of merit for the quantumness of a quantum multiparameter estimation model~\cite{Carollo2019a}.
In addition, since \(\widetilde c_{jk}\) is better than \(c_{jk}\) to manifest the regrets tradeoff for mixed states, it may be possible to take the quantity \(\tr|\sqrt{\rho_\theta} [L_j, L_k] \sqrt{\rho_\theta}|\) as an alternative form of quantum curvature.

\begin{acknowledgments}
We thank A.~Carollo for the useful discussions and for making us aware of Ref.~\cite{Carollo2019a}. 
This work is supported by 
the National Natural Science Foundation of China (Grants No.~61871162, No.~11805048, No.~11935012, and No.~11875231)
and Zhejiang Provincial Natural Science Foundation of China (Grant No.~LY18A050003).
\end{acknowledgments}

\onecolumngrid
\newpage

\setcounter{equation}{0}
\setcounter{figure}{0}
\setcounter{table}{0}
\makeatletter
\renewcommand{\theequation}{S\arabic{equation}}
\renewcommand{\thefigure}{S\arabic{figure}}
\renewcommand{\thetable}{S\arabic{table}}


\begin{center}
{\huge{}Supplemental material}
\end{center}



\section{Detailed derivation of the regret tradeoff relation \label{sec:tradeoff}}

\subsection{Regret of Fisher information}
Firstly, let \( \hs_\mathrm{s} \) be the Hilbert space associated with the underlying quantum system, and \( \mathcal{S}(\hs) \) the set of all density operators on a Hilbert space \(\hs\).
For a given POVM \(M\) acting on \(\hs_\mathrm{s}\), define the measurement channel \( \Phi:\mathcal{S}(\hs_\mathrm{s}) \to \mathcal{S}(\hs_\mathrm{r}) \) by
\begin{equation}
	\Phi: \rho \mapsto \sum_x \tr(\povm_x\rho) \dyad x,
\end{equation}
where \( \hs_\mathrm{r} \) is the Hilbert space for a register that associates the outcomes \(x\) with an orthonormal basis \( \{\ket x\} \).
The output density operators \( \Phi(\rho_\theta) \) are diagonal with the basis \( \{\ket x\} \) and its SLD operator with respect to \( \theta_j \) can be expressed as
\begin{equation}
	\widetilde L_j  = \pdv{\ln \tr(\povm_x \rho_\theta)}{\theta_j} \dyad x.
\end{equation}
Denote by \(\widetilde \qfi\) the quantum Fisher information matrix of \(\Phi(\rho_\theta)\).
Note that \(\widetilde \qfi\) equals the classical Fisher information matrix under the measurement \(M\), i.e., \(\widetilde \qfi = F(M)\).

Secondly, it can be shown that
\begin{align}
	\widetilde \qfi_{jk} 
	= \Re \tr[\widetilde L_j \widetilde L_k \Phi(\rho_\theta)] 
	= \tr[\widetilde L_j \frac{\widetilde L_k \Phi(\rho_\theta) + \Phi(\rho_\theta) \widetilde L_k}{2} ] 
	= \tr[\widetilde L_j \frac{\partial \Phi(\rho_\theta)}{\partial \theta_k}] 
	= \tr[\widetilde L_j \Phi\left( \frac{\partial \rho_\theta}{\partial \theta_k} \right)],
\end{align}
where we have used the SLD equation 
\begin{equation}
	\frac{\partial\Phi(\rho_\theta)}{\partial \theta_j} = \frac12 \qty[\widetilde L_j \Phi(\rho_\theta) + \Phi(\rho_\theta)\widetilde L_j]
\end{equation}
in the third equality and \( \partial \Phi(\rho_\theta) / \partial \theta_j = \Phi(\partial \rho_\theta / \partial \theta_j)\) in the fourth equality.
Now, let us introduce the dual map \(\Phi^\dagger\) that satisfies 
\begin{equation}
	\tr[\Phi^\dagger(X)\rho] = \tr[X\Phi(\rho)]
\end{equation}
for any density operator \( \rho \) on \(\hs_\mathrm{s}\) and any bounded operator \( X \) on \(\hs_\mathrm{r}\).
It then follows that
\begin{align}
	\widetilde \qfi_{jk} 
	& = \tr[\Phi^\dagger(\widetilde L_j) \frac{\partial \rho_\theta}{\partial \theta_k}] 
	  = \Re \tr \left( \Phi^\dagger(\widetilde L_j) L_k \rho_\theta \right), \label{eq:ljk}
\end{align}
where we have used the SLD equation \( \partial \rho_\theta / \partial \theta_k = (L_k \rho_\theta + \rho_\theta L_k)/2 \). 
Since the quantum Fisher information matrix is symmetric, by interchanging the subscripts \(j\) and \(k\) in the right hand side of Eq.~\eqref{eq:ljk}, we can also get 
\begin{equation}
	\widetilde \qfi_{jk} = \Re \tr[\Phi^\dagger(\widetilde L_k) L_j \rho_\theta].
\end{equation}
For the simplicity of notation, let us define \(\ev{\bullet}:=\tr(\bullet\rho_\theta)\).
By noting that 
\begin{align*}
	\widetilde \qfi_{jk} 
	= \Re\ev*{L_j \Phi^\dagger (\widetilde L_k)} 
	= \Re\ev*{\Phi^\dagger (\widetilde L_j) L_k} 
	= \Re\ev*{\Phi^\dagger (\widetilde L_j \widetilde L_k)},
\end{align*}
we show that the entries of the regret matrix can be expressed as
\begin{align}
	R_{jk} 	:= \qfi_{jk} - \widetilde \qfi_{jk} 
    =\Re \ev*{
    	L_j L_k
    	- L_j \Phi^\dagger (\widetilde L_k)
    	- \Phi^\dagger (\widetilde L_j) L_k
    	+ \Phi^\dagger (\widetilde L_j \widetilde L_k)
    }.\label{eq:proof2}
\end{align}

Thirdly, we shall show that the expression Eq.~\eqref{eq:proof2} of the regret matrix entries can be rewritten in a more elegant way through the open-system representation of quantum channels.
There always exist a unitary operator \(U\) on \(\hs_\mathrm{s} \otimes \hs_\mathrm{r} \otimes \hs_\mathrm{r}\) such that 
\begin{align}
	\Phi(\rho) = \tr_{1,3}[U (\rho \otimes \dyad0 \otimes \dyad0) U^\dagger]
\end{align}
for all density operators \(\rho\) on \(\hs_\mathrm{s}\), where \(\ket0\) can be any state in \(\hs_\mathrm{r}\) and \(\tr_{1,3}\) is the partial trace over the first and third tensor factors of $\hs_\mathrm{s} \otimes \hs_\mathrm{r} \otimes \hs_\mathrm{r}$ (see Ref.~\cite[Chapter 2]{WolfBook}).
With this open-system representation of \(\Phi\), it can be shown that
\begin{align}
	\tr[X \Phi (\rho)] = \tr[(\openone_\mathrm{s}\otimes X\otimes\openone_\mathrm{r}) U(\rho\otimes\dyad0 \otimes \dyad0)U^\dagger],
\end{align}
implying that
\begin{equation}
	\Phi^\dagger(X) = \tr_{2,3}\qty[
		U^\dagger (\openone_\mathrm{s} \otimes X \otimes \openone_\mathrm{r}) U 
		(\openone_\mathrm{s}\otimes \dyad0 \otimes \dyad0)
	].
\end{equation}
With this open-system representation of the dual map of \(\Phi\), Eq.~\eqref{eq:proof2} can be expressed as
\begin{equation}\label{eq:regret}
	R_{jk} = \Re\tr[ N_j N_k (\rho_\theta\otimes\dyad0 \otimes \dyad0) ],
\end{equation}
where \(N_j:= \mathcal L_j - L_j \otimes \openone_\mathrm{r} \otimes \openone_\mathrm{r}\) with \(\mathcal L_j := U^\dagger (\openone_\mathrm{s} \otimes\widetilde L_j \otimes \openone_\mathrm{r}) U \).

\subsection{Measurement uncertainty relation}
We here briefly introduce the measurement uncertainty relations, which will be invoked to derive the regret tradeoff relation.
Let \(A\) and \(B\) be two Hermitian operators standing for the ideal observables we intend to measure.
In quantum mechanics, when \([A,B] \neq 0\), these two observables cannot be jointly measured.
To approximate the joint measurement of \(A\) and \(B\) when \([A,B] \neq 0\), we can measure another pair of commuting observables \(\mathcal A\) and \(\mathcal B\) acting on the system possibly dilated by adding an ancilla whose state is denoted by a density operator \(\eta\)~\cite{Arthurs1965}.
Ozawa proposed to quantify the (state-dependent) measurement errors for the ideal observables \(A\) and \(B\) in the quantum state \(\rho\) by 
\begin{equation}
	\epsilon_A = \sqrt{\tr[(\mathcal A - A \otimes \idop)^2 (\rho\otimes\eta)]}
	\quad\mbox{and}\quad
	\epsilon_B = \sqrt{\tr[(\mathcal B - B \otimes \idop)^2 (\rho\otimes\eta)]}, 
\end{equation}
respectively, and derived the following measurement uncertainty relation~\cite{Ozawa2003,Ozawa2004a}:
\begin{equation}
	\epsilon_A \epsilon_B + \epsilon_A \sigma_B + \epsilon_B \sigma_A \geq C_{AB} := \frac12 \abs{\tr([A,B]\rho)},
\end{equation}	
where \(\sigma_A := \sqrt{\tr(A^2 \rho) - \tr(A \rho)^2}\) and \(\sigma_B := \sqrt{\tr(B^2 \rho) - \tr(B \rho)^2}\). 
Branciard obtained a stronger inequality~\cite{Branciard2013}:
\begin{equation} 
    \epsilon_A^2 \sigma_B^2 + \epsilon_B^2 \sigma_A^2 
    + 2 \sqrt{\sigma_A^2 \sigma_B^2 - C_{AB}^2} \epsilon_A \epsilon_B 
    \geq C_{AB}^2,
\end{equation}
which implies Ozawa's inequality and is tight when \(\rho\) is a pure state.
For mixed states, Ozawa showed that Branciard's inequality can be strengthen by replacing \(C_{AB}\) by 
\begin{equation}
	D_{AB} := \frac12 \tr|\sqrt\rho [A,B] \sqrt\rho|
\end{equation}
with \(|X| := \sqrt{X^\dagger X}\) for an operator \(X\).

\subsection{Derivation of the regret tradeoff relation}
Now, we derive our regret tradeoff relation.
It follows from Eq.~\eqref{eq:regret} that 
\begin{equation}
	R_{jj} = \tr[(\mathcal L_j - L_j \otimes \idop_\mathrm{r} \otimes \idop_\mathrm{r})^2 (\rho\otimes \dyad0 \otimes \dyad0)],
\end{equation}
which is in the form of Ozawa's definition of measurement error by taking \(L_j \to A\), \(\mathcal L_j \to \mathcal A\), and \(\dyad0 \otimes \dyad0 \to \eta\).
Correspondingly, we have 
\begin{equation}
	\sigma_A = \sqrt{\tr(L_j^2\rho) - \tr(L_j\rho)^2} = \sqrt{\qfi_{jj}}.
\end{equation}
Let us consider another parameter \(\theta_k\) and take \(L_k \to B\) and \(\mathcal L_k \to \mathcal B\).
It is easy to see that \([\mathcal L_j,\mathcal L_k]=0\) for \([\widetilde L_j,\widetilde L_k]=0\).
Consequently, the square roots of the regret of Fisher information for different parameters \( \theta_j \) and \(\theta_k\), i.e., \(\sqrt{R_{jj}}\) and \(\sqrt{R_{kk}}\), can be considered as the measurement errors of measuring a pair of commuting observable \(\mathcal L_j\) and \(\mathcal L_k\) in a dilated system to approximate the measurement of \( L_j \) and \( L_k \), which are unable to be jointly measured for the case of \([L_j,L_k]\neq0\).
It then follows from Branciard's inequality that 
\begin{equation}
	R_{jj} \qfi_{kk} + R_{kk} \qfi_{jj} + 2 \sqrt{
		\qfi_{jj} \qfi_{kk} - C_{jk}^2 
	} \sqrt{R_{jj} R_{kk}}
	\geq C_{jk}^2,
\end{equation}
where \(C_{jk} := \frac12 |\tr([L_j,L_k]\rho)|\).
Dividing both sides of the above inequality by \(\qfi_{jj} \qfi_{kk}\), we get our regret tradeoff relation in the main text.

\section{Estimating coherent state} \label{sec:coherent}
Here, we give the detailed calculations for the first example in the main text, i.e., the joint estimation of parameters \(\theta_1 = \Re\alpha\) and \(\theta_2 = \Im\alpha\) in coherent states \(\ket\alpha\).
Due to
\begin{align}
    \pdv{\ket\alpha}{\theta_1} = (-\theta_1 + a^\dagger) \ket\psi,
    \qq{}
    \pdv{\ket\alpha}{\theta_2} = (-\theta_2 + i a^\dagger) \ket\psi
\end{align}
and 
\begin{equation}
	\mathcal Q_{jk} = 4 \qty(\pdv{\bra\alpha}{\theta_j}) 
	\qty(\idop-\dyad{\alpha})
	\qty(\pdv{\ket\alpha}{\theta_k}),
\end{equation}
it can be shown that
\begin{equation}
	\mathcal Q =4\mqty(1 & i \\ -i & 1).
\end{equation}
We then get the regret tradeoff relation \(R_{11} + R_{22} \geq 4\), which is equivalent to \(F_{11} + F_{22} \leq 4\).

Following Helstrom~\cite{Helstrom1973a} and Holevo~\cite{Holevo1982}, we consider the measurement for jointly estimating the two parameters \(\theta_1\) and \(\theta_2\) by using an ancillary mode whose annihilation operator is denoted by \(a'\) and satisfies \([a',a^\dagger]=0\).
Define the dimensionless coordinate and momentum operators for these two modes by
\begin{align}
	Q  &= \frac{a + a^\dagger}{2}, 		\qq{}  P = \frac{a - a^\dagger}{2i}, \nonumber\\ 
	Q' &= \frac{a' + a'^\dagger}{2}, 	\qq{}  P = \frac{a' - a'^\dagger}{2i}. 
\end{align}
Note that \([Q, P] = i/2\) due to \([a,a^\dagger]=1\).
The two observables \(\mathcal A = Q - Q'\) and \( \mathcal B = P + P'\) can be jointly measured, as they are commuting.
It can be shown that \(\ev{\mathcal A} = \theta_1\), \(\ev{\mathcal B} = \theta_2\), \(\Delta \mathcal A^2 = \Delta Q^2 + \Delta Q'^2 = 1/2\), and \(\Delta \mathcal B^2 = \Delta P^2 + \Delta P'^2 = 1/2\).

To calculate the Fisher information matrix under the joint measurement of \( Q - Q'\) and \(P + P'\), we need obtain the joint probability density function with respect to the corresponding outcomes, which are denoted by \(\xi\) and \(\eta\).
Denote by \(\ket{\xi,\eta}\) the simultaneous eigenstates of the commuting observables \(Q - Q'\) and \(P + P'\). 
It is known that~\cite{Helstrom1973a}
\begin{align}
	\ket{\xi,\eta} &= \pi^{-1/2} \int e^{2 i \eta q} \ket{q}_Q \otimes \ket{q - \xi}_{Q'} \dd{q},
\end{align}
where \(\ket{q}_Q\) and \(\ket{q-x}_{Q'}\) are the eigenstates of \(Q\) and \(Q'\) with the eigenvalues \(q\) and \(q-\xi\), respectively.
Note that \(\ket{\xi,\eta}\) are normalized so that \(\ip{\xi',\eta'}{\xi,\eta} = \delta(\xi-\xi') \delta(\eta-\eta')\).
According to Born's rule in quantum mechanics, the joint probability density function of the outcomes of measuring \(Q - Q'\) and \(P + P'\) is given by \(p(\xi,\eta) = |\bra{\xi,\eta} \ket{\alpha} \otimes\ket0 |^2\).
The coherent state in the coordinate representation is given by the wave function~\cite[Section 5.1.1]{Garrison2008}
\begin{equation} \label{eq:wavefun_coherent}
	\psi_\alpha(q) := \braket{q}{\alpha} = \qty(\frac2\pi)^{1/4} \exp[- (q - \theta_1)^2] \exp(2 i q \theta_2).
\end{equation}
Therefore, 
\begin{align}
	p(\xi,\eta)
	&= \frac1\pi \abs{\int e^{-2 i \eta q} \psi_\alpha(q) \psi_0(q - \xi) \dd{q}}^2 \\
	&= \frac2{\pi^2} \abs{\int e^{-2 i \eta q} e^{- (q - \theta_1)^2} e^{2 i q \theta_2} e^{-(q - \xi)^2} \dd{q}}^2 \\
	&= \frac1\pi \exp[-(\eta - \theta_2)^2 - (\xi - \theta_1)^2].
\end{align}
The classical Fisher information matrix of \(p(\xi,\eta)\) is 
\begin{equation}
	F = \mqty(2 & 0 \\ 0 & 2),
\end{equation}
which saturates the tradeoff \(F_{11} + F_{22} \leq 4\).

Now, we consider the joint measurement of \(Q - Q'\) and \(P + P'\) with the ancillary mode being in the squeezed state \(S(\zeta)\ket0\), where \(S(\zeta)=\exp[(\zeta^* a'^2 -\zeta a'^{\dagger2})/2]\) with \(\zeta = re^{i\varphi}\) being an arbitrary complex number is the squeeze operator.
Here, we set \(\varphi = 0\).
It then can be shown that
\begin{align}
	S(\zeta)^\dagger a' S(\zeta) &= a' \cosh r - a'^\dagger \sinh r,
\end{align}
implying that
\begin{align}
	S(\zeta)^\dagger Q' S(\zeta) = Q' e^{-r}
	\qand 
	S(\zeta)^\dagger P' S(\zeta) = P' e^{r}.
\end{align}
Therefore, jointly measuring \(Q-Q'\) and \(P+P'\) with the ancillary mode being in the squeezed state \(S(\zeta)\ket0\) is equivalent to jointly measuring \(\mathcal A_r := Q - e^{-r} Q'\) and \(\mathcal B_r := P + e^{r} P'\) with the ancillary mode being in the vacuum state.

The normalized simultaneous eigenstates of \(\mathcal A_r\) and \(\mathcal B_r\) are given by 
\begin{equation}
	\ket{\xi,\eta}_r = \frac{e^{r/2}}{\sqrt{\pi}} \int e^{2i\eta q} \ket{q}_Q \otimes \ket{e^r (q-\xi)}_{Q'} \dd{q}. 
\end{equation}
It is easy to see that \(\mathcal A_r \ket{\xi, \eta}_r = \xi \ket{\xi, \eta}_r\) and \(\ip{\xi',\eta'}{\xi,\eta}_r = \delta(\xi-\xi') \delta(\eta-\eta')\). 
To show that \(\ket{\xi, \eta}_r\) is the eigenstate of \(\mathcal B_r\) with the eigenvalue \(\eta\), we need to write \(\ket{\xi,\eta}_r\) with the momentum representation.
Using \(\prescript{}{P}{\ip{p}{q}}_Q=\frac{1}{\sqrt{\pi}}e^{-2ipq}\) with \(\ket{p}_P\) denoting the eigenstate of \(P\), we get
\begin{align}
	\ket{\xi,\eta}_r &= \frac{e^{r/2}}{\pi^{3/2}} \iiint \exp[2 i \eta q - 2 i p_1 q - 2 i p_2 e^r (q - \xi)]
		\ket{p_1}_P \otimes \ket{p_2}_{P'} \dd{q} \dd{p_1} \dd{p_2} \\
	&= \frac{e^{r/2}}{\sqrt\pi} \iint \delta(\eta -  p_1 - p_2 e^r) \exp(2 i p_2 \xi e^r) 
		\ket{p_1}_P \otimes \ket{p_2}_{P'} \dd{p_1} \dd{p_2} \\
	&= \frac{e^{-r/2}}{\sqrt\pi} \int \exp[2 i \xi (\eta -  p)] 
		\ket{p}_P \otimes \ket{e^{-r}(\eta - p)}_{P'} \dd{p}. \label{eq:momentum_rep}
\end{align}
With Eq.~\eqref{eq:momentum_rep}, it is easy to see that \(\mathcal B_r \ket{\xi,\eta}_r = \eta\ket{\xi,\eta}_r\).

To calculate the classical Fisher information matrix, we need the joint probability density function:
\begin{align}
	p(\xi, \eta) = \frac{e^r}\pi \abs{\int e^{-2i\eta q} \psi_\alpha(q) \psi_0(e^r (q-\xi)) \dd{q}}^2
\end{align}
With the wave function of coherent state, namely, Eq.~\eqref{eq:wavefun_coherent}, we get 
\begin{align}
	p(\xi, \eta) &= \frac{2 e^r}{\pi^2} \abs{\int \exp[-2i\eta q - (q - \theta_1)^2 + 2 i q \theta_2 - e^{2r} (q - \xi)^2 ] \dd{q}}^2 \\
	& = \frac1{\pi} \frac{2e^r}{e^{2r} + 1} \exp[-\frac{2 (\eta -\theta_2)^2}{e^{2 r}+1} - \frac{2 e^{2 r} (\xi-\theta_1)^2}{e^{2 r}+1}]
\end{align}
This probability density function is Gaussian with the covariance matrix as follows: 
\begin{equation}
	\Sigma = \mqty( 
		\frac{e^{2 r} + 1}{4} & 0 \\ 
		0 & \frac{e^{2 r} + 1}{4 e^{2 r}}
	).
\end{equation}
The classical Fisher information matrix with respect to \(\theta_1\) and \(\theta_2\) is then given by 
\begin{equation}
	F = \Sigma^{-1} = \mqty( 
		\frac{4}{e^{2 r} + 1} & 0 \\ 
		0 & \frac{4 e^{2 r}}{e^{2 r} + 1}
	),
\end{equation}
which saturates the tradeoff relation \(F_{11} + F_{22} \leq 4\).

\section{Joint estimation of phase and phase diffusion} \label{sec:diffusion}
Here, we give the details of the calculation for the second example in the main text, i.e., the joint estimation of phase shift and phase diffusion.
The density operators of a two-level quantum system can always be represented by \(\rho = \frac12 \qty(I + \vb{r} \cdot \vb*{\sigma})\), where \(I\) is the \(2\times2\) identity matrix, \(\vb r = (r_1, r_2, r_3) \in \mathbb R^3\), and \(\vb*\sigma = (\sigma_1, \sigma_2, \sigma_3)\) is the vector of Pauli matrices.
For the joint estimation of phase and phase diffusion~\cite{Vidrighin2014}, the Bloch vector \(\vb{r}\) is given by \(\vb{r} = 	e^{-\theta_2^2} \vb{n}\), where \(\vb n = (\sin\chi \cos\theta_1, \sin\chi \sin\theta_1, \cos\chi)\) is a unit vector.
Here, \(\theta_1\) and \(\theta_2\) are the parameters of interest and \(\chi\) is a parameter determined by the initialization of the quantum system. 

To obtain the SLD operators, we use the eigenvalue decomposition of \(\rho\). 
The eigen-projections of \(\rho_\theta\) are \(\Pi_\pm = \frac12 (I \pm \vb{n} \cdot \vb*{\sigma})\) with the eigenvalues \(\lambda_\pm = [1 \pm \exp(-\theta_2^2)] / 2\).
The SLD operators about \(\theta_j\) can be expressed as
\begin{align}
	L_j &= \sum_{u,v=\pm} \frac{2}{\lambda_u + \lambda_v} \Pi_u (\partial_j \rho_\theta) \Pi_v.
	= \sum_{u,v=\pm} \frac{1}{\lambda_u + \lambda_v} \Pi_u (\partial_j \vb r \cdot \vb*\sigma) \Pi_v.
\end{align}
Substituting the expressions of \(\lambda_\pm\), \(\Pi_\pm\), and \(\vb r\) into the above formula, we get 
\begin{align}
	L_1 = \mqty(
		0 & -i e^{-\theta_2^2-i \theta_1} \\
 		i e^{-\theta_2^2+i \theta_1} & 0 
	), \qq{}
	L_2 =
	\frac1{e^{2 \theta_2^2}-1}\mqty(
 		2 \theta_2 & -2 \theta_2 e^{\theta_2^2-i \theta_1} \\
 		-2 \theta_2 e^{\theta_2^2+i \theta_1} & 2 \theta_2 
	).
\end{align}
Therefore, we get the quantum geometric tensor:
\begin{align}
	\mathcal Q = \mqty(
 		e^{-2 \theta_2^2} & 0 \\
 		0 & \frac{4 \theta_2^2}{e^{2 \theta_2^2}-1} 
	),
\end{align}
from which we can see \(c_{12}=0\).
To calculate \(\widetilde c_{12}\), we note that
\begin{align}
	[L_1, L_2] = \frac{-i}{e^{2 \theta_2^2}-1}\mqty(
 		-4 \theta_2 & 0 \\
 		0 & 4 \theta_2 
	).
\end{align}
After some algebras, we get 
\begin{equation}
	\tr|\sqrt\rho [L_1, L_2] \sqrt\rho|
	=\frac{4 e^{-\theta_2^2} \theta_2}{\sqrt{e^{2 \theta_2^2}-1}}
\end{equation}
and thus
\begin{equation}
	\widetilde c_{12} = \frac{\tr|\sqrt\rho [L_1, L_2] \sqrt\rho|}{2\sqrt{\Re Q_{11} \Re Q_{22}}} = 1.
\end{equation}


\begin{thebibliography}{72}%
\makeatletter
\providecommand \@ifxundefined [1]{%
 \@ifx{#1\undefined}
}%
\providecommand \@ifnum [1]{%
 \ifnum #1\expandafter \@firstoftwo
 \else \expandafter \@secondoftwo
 \fi
}%
\providecommand \@ifx [1]{%
 \ifx #1\expandafter \@firstoftwo
 \else \expandafter \@secondoftwo
 \fi
}%
\providecommand \natexlab [1]{#1}%
\providecommand \enquote  [1]{``#1''}%
\providecommand \bibnamefont  [1]{#1}%
\providecommand \bibfnamefont [1]{#1}%
\providecommand \citenamefont [1]{#1}%
\providecommand \href@noop [0]{\@secondoftwo}%
\providecommand \href [0]{\begingroup \@sanitize@url \@href}%
\providecommand \@href[1]{\@@startlink{#1}\@@href}%
\providecommand \@@href[1]{\endgroup#1\@@endlink}%
\providecommand \@sanitize@url [0]{\catcode `\\12\catcode `\$12\catcode
  `\&12\catcode `\#12\catcode `\^12\catcode `\_12\catcode `\%12\relax}%
\providecommand \@@startlink[1]{}%
\providecommand \@@endlink[0]{}%
\providecommand \url  [0]{\begingroup\@sanitize@url \@url }%
\providecommand \@url [1]{\endgroup\@href {#1}{\urlprefix }}%
\providecommand \urlprefix  [0]{URL }%
\providecommand \Eprint [0]{\href }%
\providecommand \doibase [0]{https://doi.org/}%
\providecommand \selectlanguage [0]{\@gobble}%
\providecommand \bibinfo  [0]{\@secondoftwo}%
\providecommand \bibfield  [0]{\@secondoftwo}%
\providecommand \translation [1]{[#1]}%
\providecommand \BibitemOpen [0]{}%
\providecommand \bibitemStop [0]{}%
\providecommand \bibitemNoStop [0]{.\EOS\space}%
\providecommand \EOS [0]{\spacefactor3000\relax}%
\providecommand \BibitemShut  [1]{\csname bibitem#1\endcsname}%
\let\auto@bib@innerbib\@empty
\bibitem [{\citenamefont {Helstrom}(1967)}]{Helstrom1967}%
  \BibitemOpen
  \bibfield  {author} {\bibinfo {author} {\bibfnamefont {C.}~\bibnamefont
  {Helstrom}},\ }\bibfield  {title} {\bibinfo {title} {Minimum mean-squared
  error of estimates in quantum statistics},\ }\href
  {https://doi.org/http://dx.doi.org/10.1016/0375-9601(67)90366-0} {\bibfield
  {journal} {\bibinfo  {journal} {Phys. Lett. A}\ }\textbf {\bibinfo {volume}
  {25}},\ \bibinfo {pages} {101} (\bibinfo {year} {1967})}\BibitemShut
  {NoStop}%
\bibitem [{\citenamefont {Helstrom}(1968)}]{Helstrom1968}%
  \BibitemOpen
  \bibfield  {author} {\bibinfo {author} {\bibfnamefont {C.}~\bibnamefont
  {Helstrom}},\ }\bibfield  {title} {\bibinfo {title} {The minimum variance of
  estimates in quantum signal detection},\ }\href
  {https://doi.org/10.1109/TIT.1968.1054108} {\bibfield  {journal} {\bibinfo
  {journal} {IEEE Trans. Inform. Theory}\ }\textbf {\bibinfo {volume} {14}},\
  \bibinfo {pages} {234} (\bibinfo {year} {1968})}\BibitemShut {NoStop}%
\bibitem [{\citenamefont {Yuen}\ and\ \citenamefont {Lax}(1973)}]{Yuen1973}%
  \BibitemOpen
  \bibfield  {author} {\bibinfo {author} {\bibfnamefont {H.}~\bibnamefont
  {Yuen}}\ and\ \bibinfo {author} {\bibfnamefont {M.}~\bibnamefont {Lax}},\
  }\bibfield  {title} {\bibinfo {title} {Multiple-parameter quantum estimation
  and measurement of nonselfadjoint observables},\ }\href
  {https://doi.org/10.1109/TIT.1973.1055103} {\bibfield  {journal} {\bibinfo
  {journal} {IEEE Trans. Inform. Theory}\ }\textbf {\bibinfo {volume} {19}},\
  \bibinfo {pages} {740} (\bibinfo {year} {1973})}\BibitemShut {NoStop}%
\bibitem [{\citenamefont {Belavkin}(1976)}]{Belavkin1976}%
  \BibitemOpen
  \bibfield  {author} {\bibinfo {author} {\bibfnamefont {V.~P.}\ \bibnamefont
  {Belavkin}},\ }\bibfield  {title} {\bibinfo {title} {Generalized uncertainty
  relations and efficient measurements in quantum systems},\ }\href
  {https://doi.org/10.1007/BF01032091} {\bibfield  {journal} {\bibinfo
  {journal} {Theor. Math. Phys.}\ }\textbf {\bibinfo {volume} {26}},\ \bibinfo
  {pages} {213} (\bibinfo {year} {1976})}\BibitemShut {NoStop}%
\bibitem [{\citenamefont {Helstrom}(1976)}]{Helstrom1976}%
  \BibitemOpen
  \bibfield  {author} {\bibinfo {author} {\bibfnamefont {C.~W.}\ \bibnamefont
  {Helstrom}},\ }\href@noop {} {\emph {\bibinfo {title} {Quantum Detection and
  Estimation Theory}}}\ (\bibinfo  {publisher} {Academic Press, New York},\
  \bibinfo {year} {1976})\BibitemShut {NoStop}%
\bibitem [{\citenamefont {Holevo}(1982)}]{Holevo1982}%
  \BibitemOpen
  \bibfield  {author} {\bibinfo {author} {\bibfnamefont {A.~S.}\ \bibnamefont
  {Holevo}},\ }\href@noop {} {\emph {\bibinfo {title} {Probabilistic and
  Statistical Aspects of Quantum Theory}}}\ (\bibinfo  {publisher}
  {North-Holland, Amsterdam},\ \bibinfo {year} {1982})\BibitemShut {NoStop}%
\bibitem [{\citenamefont {Personick}(1971)}]{Personick1971}%
  \BibitemOpen
  \bibfield  {author} {\bibinfo {author} {\bibfnamefont {S.}~\bibnamefont
  {Personick}},\ }\bibfield  {title} {\bibinfo {title} {Application of quantum
  estimation theory to analog communication over quantum channels},\ }\href
  {https://doi.org/10.1109/TIT.1971.1054643} {\bibfield  {journal} {\bibinfo
  {journal} {IEEE Trans. Inform. Theory}\ }\textbf {\bibinfo {volume} {17}},\
  \bibinfo {pages} {240} (\bibinfo {year} {1971})}\BibitemShut {NoStop}%
\bibitem [{\citenamefont {Hayashi}(2005)}]{HayashiBook}%
  \BibitemOpen
  \bibinfo {editor} {\bibfnamefont {M.}~\bibnamefont {Hayashi}},\ ed.,\ \href
  {http://gen.lib.rus.ec/book/index.php?md5=512D187517928E92C917BD6312DF38CD}
  {\emph {\bibinfo {title} {Asymptotic theory of quantum statistical inference:
  Selected Papers}}}\ (\bibinfo  {publisher} {World Scientific Publishing
  Company},\ \bibinfo {year} {2005})\BibitemShut {NoStop}%
\bibitem [{\citenamefont {Tsang}\ \emph {et~al.}(2020)\citenamefont {Tsang},
  \citenamefont {Albarelli},\ and\ \citenamefont {Datta}}]{Tsang2020a}%
  \BibitemOpen
  \bibfield  {author} {\bibinfo {author} {\bibfnamefont {M.}~\bibnamefont
  {Tsang}}, \bibinfo {author} {\bibfnamefont {F.}~\bibnamefont {Albarelli}},\
  and\ \bibinfo {author} {\bibfnamefont {A.}~\bibnamefont {Datta}},\ }\bibfield
   {title} {\bibinfo {title} {Quantum semiparametric estimation},\ }\href
  {https://doi.org/10.1103/PhysRevX.10.031023} {\bibfield  {journal} {\bibinfo
  {journal} {Phys. Rev. X}\ }\textbf {\bibinfo {volume} {10}},\ \bibinfo
  {pages} {031023} (\bibinfo {year} {2020})}\BibitemShut {NoStop}%
\bibitem [{\citenamefont {Fisher}(1922)}]{Fisher1922}%
  \BibitemOpen
  \bibfield  {author} {\bibinfo {author} {\bibfnamefont {R.~A.}\ \bibnamefont
  {Fisher}},\ }\bibfield  {title} {\bibinfo {title} {On the mathematical
  foundations of theoretical statistics},\ }\href
  {http://www.jstor.org/stable/91208} {\bibfield  {journal} {\bibinfo
  {journal} {Philosophical Transactions of the Royal Society of London. Series
  A, Containing Papers of a Mathematical or Physical Character}\ }\textbf
  {\bibinfo {volume} {222}},\ \bibinfo {pages} {309} (\bibinfo {year}
  {1922})}\BibitemShut {NoStop}%
\bibitem [{\citenamefont {Fisher}(1925)}]{Fisher1925}%
  \BibitemOpen
  \bibfield  {author} {\bibinfo {author} {\bibfnamefont {R.~A.}\ \bibnamefont
  {Fisher}},\ }\bibfield  {title} {\bibinfo {title} {Theory of statistical
  estimation},\ }\href {https://doi.org/10.1017/S0305004100009580} {\bibfield
  {journal} {\bibinfo  {journal} {Mathematical Proceedings of the Cambridge
  Philosophical Society}\ }\textbf {\bibinfo {volume} {22}},\ \bibinfo {pages}
  {700} (\bibinfo {year} {1925})}\BibitemShut {NoStop}%
\bibitem [{\citenamefont {Cram\'{e}r}(1946)}]{Cramer1946}%
  \BibitemOpen
  \bibfield  {author} {\bibinfo {author} {\bibfnamefont {H.}~\bibnamefont
  {Cram\'{e}r}},\ }\href@noop {} {\emph {\bibinfo {title} {Mathematical Methods
  of Statistics}}}\ (\bibinfo  {publisher} {Princeton University Press,
  Princeton},\ \bibinfo {year} {1946})\BibitemShut {NoStop}%
\bibitem [{\citenamefont {Rao}(1945)}]{Rao1945}%
  \BibitemOpen
  \bibfield  {author} {\bibinfo {author} {\bibfnamefont {C.~R.}\ \bibnamefont
  {Rao}},\ }\bibfield  {title} {\bibinfo {title} {Information and the accuracy
  attainable in the estimation of statistical parameters},\ }\bibfield
  {booktitle} {\emph {\bibinfo {booktitle} {Breakthroughs in Statistics:
  Foundations and Basic Theory}},\ }\href
  {https://doi.org/10.1007/978-1-4612-0919-5_16} {\bibfield  {journal}
  {\bibinfo  {journal} {Bull. Calcutta Math. Soc.}\ }\textbf {\bibinfo {volume}
  {37}},\ \bibinfo {pages} {81} (\bibinfo {year} {1945})}\BibitemShut {NoStop}%
\bibitem [{\citenamefont {Kay}(1993)}]{Kay1993}%
  \BibitemOpen
  \bibfield  {author} {\bibinfo {author} {\bibfnamefont {S.~M.}\ \bibnamefont
  {Kay}},\ }\href
  {http://gen.lib.rus.ec/book/index.php?md5=130E67A3B79A59BFC6C759EA771F1D50}
  {\emph {\bibinfo {title} {Fundamentals of Statistical Signal Processing,
  Volume I: Estimation Theory}}}\ (\bibinfo  {publisher} {Prentice Hall},\
  \bibinfo {year} {1993})\BibitemShut {NoStop}%
\bibitem [{\citenamefont {Wasserman}(2010)}]{Wasserman2010}%
  \BibitemOpen
  \bibfield  {author} {\bibinfo {author} {\bibfnamefont {L.}~\bibnamefont
  {Wasserman}},\ }\href@noop {} {\emph {\bibinfo {title} {All of Statistics: A
  Concise Course in Statistical Inference}}}\ (\bibinfo  {publisher} {Springer
  Publishing Company, Incorporated},\ \bibinfo {year} {2010})\BibitemShut
  {NoStop}%
\bibitem [{\citenamefont {Casella}\ and\ \citenamefont
  {Berger}(2002)}]{Casella2002}%
  \BibitemOpen
  \bibfield  {author} {\bibinfo {author} {\bibfnamefont {G.}~\bibnamefont
  {Casella}}\ and\ \bibinfo {author} {\bibfnamefont {R.~L.}\ \bibnamefont
  {Berger}},\ }\href@noop {} {\emph {\bibinfo {title} {Statistical
  Inference}}},\ \bibinfo {edition} {2nd}\ ed.\ (\bibinfo  {publisher} {Duxbury
  Press, Pacific Grove},\ \bibinfo {year} {2002})\BibitemShut {NoStop}%
\bibitem [{\citenamefont {Lehmann}\ and\ \citenamefont
  {Casella}(1998)}]{Lehmann1998}%
  \BibitemOpen
  \bibfield  {author} {\bibinfo {author} {\bibfnamefont {E.}~\bibnamefont
  {Lehmann}}\ and\ \bibinfo {author} {\bibfnamefont {G.}~\bibnamefont
  {Casella}},\ }\href@noop {} {\emph {\bibinfo {title} {Theory of Point
  Estimation}}}\ (\bibinfo  {publisher} {Springer-Verlag, New York},\ \bibinfo
  {year} {1998})\BibitemShut {NoStop}%
\bibitem [{\citenamefont {Braunstein}\ and\ \citenamefont
  {Caves}(1994)}]{Braunstein1994}%
  \BibitemOpen
  \bibfield  {author} {\bibinfo {author} {\bibfnamefont {S.~L.}\ \bibnamefont
  {Braunstein}}\ and\ \bibinfo {author} {\bibfnamefont {C.~M.}\ \bibnamefont
  {Caves}},\ }\bibfield  {title} {\bibinfo {title} {Statistical distance and
  the geometry of quantum states},\ }\href
  {https://doi.org/10.1103/PhysRevLett.72.3439} {\bibfield  {journal} {\bibinfo
   {journal} {Phys. Rev. Lett.}\ }\textbf {\bibinfo {volume} {72}},\ \bibinfo
  {pages} {3439} (\bibinfo {year} {1994})}\BibitemShut {NoStop}%
\bibitem [{\citenamefont {Braunstein}\ \emph {et~al.}(1996)\citenamefont
  {Braunstein}, \citenamefont {Caves},\ and\ \citenamefont
  {Milburn}}]{Braunstein1996}%
  \BibitemOpen
  \bibfield  {author} {\bibinfo {author} {\bibfnamefont {S.~L.}\ \bibnamefont
  {Braunstein}}, \bibinfo {author} {\bibfnamefont {C.~M.}\ \bibnamefont
  {Caves}},\ and\ \bibinfo {author} {\bibfnamefont {G.}~\bibnamefont
  {Milburn}},\ }\bibfield  {title} {\bibinfo {title} {Generalized uncertainty
  relations: Theory, examples, and lorentz invariance},\ }\href
  {https://doi.org/10.1006/aphy.1996.0040} {\bibfield  {journal} {\bibinfo
  {journal} {Ann. Phys.}\ }\textbf {\bibinfo {volume} {247}},\ \bibinfo {pages}
  {135 } (\bibinfo {year} {1996})}\BibitemShut {NoStop}%
\bibitem [{\citenamefont {Fujiwara}(2006)}]{Fujiwara2006}%
  \BibitemOpen
  \bibfield  {author} {\bibinfo {author} {\bibfnamefont {A.}~\bibnamefont
  {Fujiwara}},\ }\bibfield  {title} {\bibinfo {title} {Strong consistency and
  asymptotic efficiency for adaptive quantum estimation problems},\ }\href
  {http://stacks.iop.org/0305-4470/39/i=40/a=014} {\bibfield  {journal}
  {\bibinfo  {journal} {J. Phys. A: Math. Gen.}\ }\textbf {\bibinfo {volume}
  {39}},\ \bibinfo {pages} {12489} (\bibinfo {year} {2006})}\BibitemShut
  {NoStop}%
\bibitem [{\citenamefont {Heisenberg}(1927)}]{Heisenberg1927}%
  \BibitemOpen
  \bibfield  {author} {\bibinfo {author} {\bibfnamefont {W.}~\bibnamefont
  {Heisenberg}},\ }\bibfield  {title} {\bibinfo {title} {The physical content
  of quantum kinematics and mechanics},\ }\href
  {https://doi.org/10.1007/BF01397280} {\bibfield  {journal} {\bibinfo
  {journal} {Z. Phys.}\ }\textbf {\bibinfo {volume} {43}},\ \bibinfo {pages}
  {172} (\bibinfo {year} {1927})},\ \bibinfo {note} {{English translation in
  \textit{Quantum Theory and Measurement}, edited by J. A. Wheeler and W. H.
  Zurek (Princeton Univ. Press, Princeton, NJ, 1984), p. 62.}}\BibitemShut
  {Stop}%
\bibitem [{\citenamefont {Busch}\ \emph {et~al.}(2007)\citenamefont {Busch},
  \citenamefont {Heinonen},\ and\ \citenamefont {Lahti}}]{Busch2007}%
  \BibitemOpen
  \bibfield  {author} {\bibinfo {author} {\bibfnamefont {P.}~\bibnamefont
  {Busch}}, \bibinfo {author} {\bibfnamefont {T.}~\bibnamefont {Heinonen}},\
  and\ \bibinfo {author} {\bibfnamefont {P.}~\bibnamefont {Lahti}},\ }\bibfield
   {title} {\bibinfo {title} {{Heisenberg}'s uncertainty principle},\ }\href
  {https://doi.org/10.1016/j.physrep.2007.05.006} {\bibfield  {journal}
  {\bibinfo  {journal} {Phys. Rep.}\ }\textbf {\bibinfo {volume} {452}},\
  \bibinfo {pages} {155 } (\bibinfo {year} {2007})}\BibitemShut {NoStop}%
\bibitem [{\citenamefont {Tsang}\ \emph {et~al.}(2016)\citenamefont {Tsang},
  \citenamefont {Nair},\ and\ \citenamefont {Lu}}]{Tsang2016b}%
  \BibitemOpen
  \bibfield  {author} {\bibinfo {author} {\bibfnamefont {M.}~\bibnamefont
  {Tsang}}, \bibinfo {author} {\bibfnamefont {R.}~\bibnamefont {Nair}},\ and\
  \bibinfo {author} {\bibfnamefont {X.-M.}\ \bibnamefont {Lu}},\ }\bibfield
  {title} {\bibinfo {title} {Quantum theory of superresolution for two
  incoherent optical point sources},\ }\href
  {https://doi.org/10.1103/PhysRevX.6.031033} {\bibfield  {journal} {\bibinfo
  {journal} {Phys. Rev. X}\ }\textbf {\bibinfo {volume} {6}},\ \bibinfo {pages}
  {031033} (\bibinfo {year} {2016})}\BibitemShut {NoStop}%
\bibitem [{\citenamefont {Tsang}(2019)}]{Tsang2019a}%
  \BibitemOpen
  \bibfield  {author} {\bibinfo {author} {\bibfnamefont {M.}~\bibnamefont
  {Tsang}},\ }\bibfield  {title} {\bibinfo {title} {Resolving starlight: a
  quantum perspective},\ }\href {https://doi.org/10.1080/00107514.2020.1736375}
  {\bibfield  {journal} {\bibinfo  {journal} {Contemp. Phys.}\ }\textbf
  {\bibinfo {volume} {60}},\ \bibinfo {pages} {279} (\bibinfo {year}
  {2019})}\BibitemShut {NoStop}%
\bibitem [{\citenamefont {Baumgratz}\ and\ \citenamefont
  {Datta}(2016)}]{Baumgratz2016}%
  \BibitemOpen
  \bibfield  {author} {\bibinfo {author} {\bibfnamefont {T.}~\bibnamefont
  {Baumgratz}}\ and\ \bibinfo {author} {\bibfnamefont {A.}~\bibnamefont
  {Datta}},\ }\bibfield  {title} {\bibinfo {title} {Quantum enhanced estimation
  of a multidimensional field},\ }\href
  {https://doi.org/10.1103/PhysRevLett.116.030801} {\bibfield  {journal}
  {\bibinfo  {journal} {Phys. Rev. Lett.}\ }\textbf {\bibinfo {volume} {116}},\
  \bibinfo {pages} {030801} (\bibinfo {year} {2016})}\BibitemShut {NoStop}%
\bibitem [{\citenamefont {Hou}\ \emph {et~al.}(2020)\citenamefont {Hou},
  \citenamefont {Zhang}, \citenamefont {Xiang}, \citenamefont {Li},
  \citenamefont {Guo}, \citenamefont {Chen}, \citenamefont {Liu},\ and\
  \citenamefont {Yuan}}]{Hou2020}%
  \BibitemOpen
  \bibfield  {author} {\bibinfo {author} {\bibfnamefont {Z.}~\bibnamefont
  {Hou}}, \bibinfo {author} {\bibfnamefont {Z.}~\bibnamefont {Zhang}}, \bibinfo
  {author} {\bibfnamefont {G.-Y.}\ \bibnamefont {Xiang}}, \bibinfo {author}
  {\bibfnamefont {C.-F.}\ \bibnamefont {Li}}, \bibinfo {author} {\bibfnamefont
  {G.-C.}\ \bibnamefont {Guo}}, \bibinfo {author} {\bibfnamefont
  {H.}~\bibnamefont {Chen}}, \bibinfo {author} {\bibfnamefont {L.}~\bibnamefont
  {Liu}},\ and\ \bibinfo {author} {\bibfnamefont {H.}~\bibnamefont {Yuan}},\
  }\bibfield  {title} {\bibinfo {title} {Minimal tradeoff and ultimate
  precision limit of multiparameter quantum magnetometry under the parallel
  scheme},\ }\href {https://doi.org/10.1103/PhysRevLett.125.020501} {\bibfield
  {journal} {\bibinfo  {journal} {Phys. Rev. Lett.}\ }\textbf {\bibinfo
  {volume} {125}},\ \bibinfo {pages} {020501} (\bibinfo {year}
  {2020})}\BibitemShut {NoStop}%
\bibitem [{\citenamefont {Vidrighin}\ \emph {et~al.}(2014)\citenamefont
  {Vidrighin}, \citenamefont {Donati}, \citenamefont {Genoni}, \citenamefont
  {Jin}, \citenamefont {Kolthammer}, \citenamefont {Kim}, \citenamefont
  {Datta}, \citenamefont {Barbieri},\ and\ \citenamefont
  {Walmsley}}]{Vidrighin2014}%
  \BibitemOpen
  \bibfield  {author} {\bibinfo {author} {\bibfnamefont {M.~D.}\ \bibnamefont
  {Vidrighin}}, \bibinfo {author} {\bibfnamefont {G.}~\bibnamefont {Donati}},
  \bibinfo {author} {\bibfnamefont {M.~G.}\ \bibnamefont {Genoni}}, \bibinfo
  {author} {\bibfnamefont {X.-M.}\ \bibnamefont {Jin}}, \bibinfo {author}
  {\bibfnamefont {W.~S.}\ \bibnamefont {Kolthammer}}, \bibinfo {author}
  {\bibfnamefont {M.}~\bibnamefont {Kim}}, \bibinfo {author} {\bibfnamefont
  {A.}~\bibnamefont {Datta}}, \bibinfo {author} {\bibfnamefont
  {M.}~\bibnamefont {Barbieri}},\ and\ \bibinfo {author} {\bibfnamefont
  {I.~A.}\ \bibnamefont {Walmsley}},\ }\bibfield  {title} {\bibinfo {title}
  {Joint estimation of phase and phase diffusion for quantum metrology},\
  }\href {http://dx.doi.org/10.1038/ncomms4532} {\bibfield  {journal} {\bibinfo
   {journal} {Nat. Commun.}\ }\textbf {\bibinfo {volume} {5}},\ \bibinfo
  {pages} {3532} (\bibinfo {year} {2014})}\BibitemShut {NoStop}%
\bibitem [{\citenamefont {Carollo}\ \emph {et~al.}(2019)\citenamefont
  {Carollo}, \citenamefont {Spagnolo}, \citenamefont {Dubkov},\ and\
  \citenamefont {Valenti}}]{Carollo2019a}%
  \BibitemOpen
  \bibfield  {author} {\bibinfo {author} {\bibfnamefont {A.}~\bibnamefont
  {Carollo}}, \bibinfo {author} {\bibfnamefont {B.}~\bibnamefont {Spagnolo}},
  \bibinfo {author} {\bibfnamefont {A.~A.}\ \bibnamefont {Dubkov}},\ and\
  \bibinfo {author} {\bibfnamefont {D.}~\bibnamefont {Valenti}},\ }\bibfield
  {title} {\bibinfo {title} {On quantumness in multi-parameter quantum
  estimation},\ }\href {https://doi.org/10.1088/1742-5468/ab3ccb} {\bibfield
  {journal} {\bibinfo  {journal} {J. Stat. Mech: Theory Exp.}\ }\textbf
  {\bibinfo {volume} {2019}},\ \bibinfo {pages} {094010} (\bibinfo {year}
  {2019})}\BibitemShut {NoStop}%
\bibitem [{\citenamefont {Rubio}\ \emph {et~al.}(2018)\citenamefont {Rubio},
  \citenamefont {Knott},\ and\ \citenamefont {Dunningham}}]{Rubio2018}%
  \BibitemOpen
  \bibfield  {author} {\bibinfo {author} {\bibfnamefont {J.}~\bibnamefont
  {Rubio}}, \bibinfo {author} {\bibfnamefont {P.}~\bibnamefont {Knott}},\ and\
  \bibinfo {author} {\bibfnamefont {J.}~\bibnamefont {Dunningham}},\ }\bibfield
   {title} {\bibinfo {title} {Non-asymptotic analysis of quantum metrology
  protocols beyond the {Cram{\'{e}}r{\textendash}Rao} bound},\ }\href
  {https://doi.org/10.1088/2399-6528/aaa234} {\bibfield  {journal} {\bibinfo
  {journal} {Journal of Physics Communications}\ }\textbf {\bibinfo {volume}
  {2}},\ \bibinfo {pages} {015027} (\bibinfo {year} {2018})}\BibitemShut
  {NoStop}%
\bibitem [{\citenamefont {Albarelli}\ \emph {et~al.}(2019)\citenamefont
  {Albarelli}, \citenamefont {Friel},\ and\ \citenamefont
  {Datta}}]{Albarelli2019}%
  \BibitemOpen
  \bibfield  {author} {\bibinfo {author} {\bibfnamefont {F.}~\bibnamefont
  {Albarelli}}, \bibinfo {author} {\bibfnamefont {J.~F.}\ \bibnamefont
  {Friel}},\ and\ \bibinfo {author} {\bibfnamefont {A.}~\bibnamefont {Datta}},\
  }\bibfield  {title} {\bibinfo {title} {Evaluating the {Holevo Cram\'er-Rao}
  bound for multiparameter quantum metrology},\ }\href
  {https://doi.org/10.1103/PhysRevLett.123.200503} {\bibfield  {journal}
  {\bibinfo  {journal} {Phys. Rev. Lett.}\ }\textbf {\bibinfo {volume} {123}},\
  \bibinfo {pages} {200503} (\bibinfo {year} {2019})}\BibitemShut {NoStop}%
\bibitem [{\citenamefont {Tsang}()}]{Tsang2019f}%
  \BibitemOpen
  \bibfield  {author} {\bibinfo {author} {\bibfnamefont {M.}~\bibnamefont
  {Tsang}},\ }\bibfield  {title} {\bibinfo {title} {The {Holevo Cram\'er-Rao}
  bound is at most thrice the {Helstrom} version},\ }\href@noop {} {\ }\Eprint
  {https://arxiv.org/abs/1911.08359} {arXiv:1911.08359} \BibitemShut {NoStop}%
\bibitem [{\citenamefont {Albarelli}\ \emph {et~al.}()\citenamefont
  {Albarelli}, \citenamefont {Tsang},\ and\ \citenamefont
  {Datta}}]{Albarelli2019a}%
  \BibitemOpen
  \bibfield  {author} {\bibinfo {author} {\bibfnamefont {F.}~\bibnamefont
  {Albarelli}}, \bibinfo {author} {\bibfnamefont {M.}~\bibnamefont {Tsang}},\
  and\ \bibinfo {author} {\bibfnamefont {A.}~\bibnamefont {Datta}},\ }\bibfield
   {title} {\bibinfo {title} {Upper bounds on the {Holevo Cram\'er-Rao} bound
  for multiparameter quantum parametric and semiparametric estimation},\
  }\href@noop {} {\ }\Eprint {https://arxiv.org/abs/1911.11036v1}
  {arXiv:1911.11036v1} \BibitemShut {NoStop}%
\bibitem [{\citenamefont {Sidhu}\ and\ \citenamefont {Kok}(2020)}]{Sidhu2020}%
  \BibitemOpen
  \bibfield  {author} {\bibinfo {author} {\bibfnamefont {J.~S.}\ \bibnamefont
  {Sidhu}}\ and\ \bibinfo {author} {\bibfnamefont {P.}~\bibnamefont {Kok}},\
  }\bibfield  {title} {\bibinfo {title} {Geometric perspective on quantum
  parameter estimation},\ }\href {https://doi.org/10.1116/1.5119961} {\bibfield
   {journal} {\bibinfo  {journal} {AVS Quantum Sci.}\ }\textbf {\bibinfo
  {volume} {2}},\ \bibinfo {pages} {014701} (\bibinfo {year}
  {2020})}\BibitemShut {NoStop}%
\bibitem [{\citenamefont {Lu}\ \emph {et~al.}(2020)\citenamefont {Lu},
  \citenamefont {Ma},\ and\ \citenamefont {Zhang}}]{Lu2020a}%
  \BibitemOpen
  \bibfield  {author} {\bibinfo {author} {\bibfnamefont {X.-M.}\ \bibnamefont
  {Lu}}, \bibinfo {author} {\bibfnamefont {Z.}~\bibnamefont {Ma}},\ and\
  \bibinfo {author} {\bibfnamefont {C.}~\bibnamefont {Zhang}},\ }\bibfield
  {title} {\bibinfo {title} {Generalized-mean {Cram\'er-Rao} bounds for
  multiparameter quantum metrology},\ }\href
  {https://doi.org/10.1103/PhysRevA.101.022303} {\bibfield  {journal} {\bibinfo
   {journal} {Phys. Rev. A}\ }\textbf {\bibinfo {volume} {101}},\ \bibinfo
  {pages} {022303} (\bibinfo {year} {2020})}\BibitemShut {NoStop}%
\bibitem [{\citenamefont {Sidhu}\ \emph {et~al.}()\citenamefont {Sidhu},
  \citenamefont {Ouyang}, \citenamefont {Campbell},\ and\ \citenamefont
  {Kok}}]{Sidhu}%
  \BibitemOpen
  \bibfield  {author} {\bibinfo {author} {\bibfnamefont {J.~S.}\ \bibnamefont
  {Sidhu}}, \bibinfo {author} {\bibfnamefont {Y.}~\bibnamefont {Ouyang}},
  \bibinfo {author} {\bibfnamefont {E.~T.}\ \bibnamefont {Campbell}},\ and\
  \bibinfo {author} {\bibfnamefont {P.}~\bibnamefont {Kok}},\ }\bibfield
  {title} {\bibinfo {title} {Tight bounds on the simultaneous estimation of
  incompatible parameters},\ }\href@noop {} {\ }\Eprint
  {https://arxiv.org/abs/1912.09218} {arXiv:1912.09218} \BibitemShut {NoStop}%
\bibitem [{\citenamefont {Ragy}\ \emph {et~al.}(2016)\citenamefont {Ragy},
  \citenamefont {Jarzyna},\ and\ \citenamefont
  {Demkowicz-Dobrza\'{n}ski}}]{Ragy2016}%
  \BibitemOpen
  \bibfield  {author} {\bibinfo {author} {\bibfnamefont {S.}~\bibnamefont
  {Ragy}}, \bibinfo {author} {\bibfnamefont {M.}~\bibnamefont {Jarzyna}},\ and\
  \bibinfo {author} {\bibfnamefont {R.}~\bibnamefont
  {Demkowicz-Dobrza\'{n}ski}},\ }\bibfield  {title} {\bibinfo {title}
  {Compatibility in multiparameter quantum metrology},\ }\href
  {https://doi.org/10.1103/PhysRevA.94.052108} {\bibfield  {journal} {\bibinfo
  {journal} {Phys. Rev. A}\ }\textbf {\bibinfo {volume} {94}},\ \bibinfo
  {pages} {052108} (\bibinfo {year} {2016})}\BibitemShut {NoStop}%
\bibitem [{\citenamefont {Li}\ \emph {et~al.}(2016)\citenamefont {Li},
  \citenamefont {Ferrie}, \citenamefont {Gross}, \citenamefont {Kalev},\ and\
  \citenamefont {Caves}}]{Li2016}%
  \BibitemOpen
  \bibfield  {author} {\bibinfo {author} {\bibfnamefont {N.}~\bibnamefont
  {Li}}, \bibinfo {author} {\bibfnamefont {C.}~\bibnamefont {Ferrie}}, \bibinfo
  {author} {\bibfnamefont {J.~A.}\ \bibnamefont {Gross}}, \bibinfo {author}
  {\bibfnamefont {A.}~\bibnamefont {Kalev}},\ and\ \bibinfo {author}
  {\bibfnamefont {C.~M.}\ \bibnamefont {Caves}},\ }\bibfield  {title} {\bibinfo
  {title} {Fisher-symmetric informationally complete measurements for pure
  states},\ }\href {https://doi.org/10.1103/PhysRevLett.116.180402} {\bibfield
  {journal} {\bibinfo  {journal} {Phys. Rev. Lett.}\ }\textbf {\bibinfo
  {volume} {116}},\ \bibinfo {pages} {180402} (\bibinfo {year}
  {2016})}\BibitemShut {NoStop}%
\bibitem [{\citenamefont {Zhu}\ and\ \citenamefont {Hayashi}(2018)}]{Zhu2018}%
  \BibitemOpen
  \bibfield  {author} {\bibinfo {author} {\bibfnamefont {H.}~\bibnamefont
  {Zhu}}\ and\ \bibinfo {author} {\bibfnamefont {M.}~\bibnamefont {Hayashi}},\
  }\bibfield  {title} {\bibinfo {title} {Universally {Fisher}-symmetric
  informationally complete measurements},\ }\href
  {https://doi.org/10.1103/PhysRevLett.120.030404} {\bibfield  {journal}
  {\bibinfo  {journal} {Phys. Rev. Lett.}\ }\textbf {\bibinfo {volume} {120}},\
  \bibinfo {pages} {030404} (\bibinfo {year} {2018})}\BibitemShut {NoStop}%
\bibitem [{\citenamefont {Suzuki}(2016)}]{Suzukia2016}%
  \BibitemOpen
  \bibfield  {author} {\bibinfo {author} {\bibfnamefont {J.}~\bibnamefont
  {Suzuki}},\ }\bibfield  {title} {\bibinfo {title} {Explicit formula for the
  {Holevo} bound for two-parameter qubit-state estimation problem},\ }\href
  {https://doi.org/10.1063/1.4945086} {\bibfield  {journal} {\bibinfo
  {journal} {J. Math. Phys.}\ }\textbf {\bibinfo {volume} {57}},\ \bibinfo
  {pages} {042201} (\bibinfo {year} {2016})}\BibitemShut {NoStop}%
\bibitem [{\citenamefont {Suzuki}(2019)}]{Suzuki2019}%
  \BibitemOpen
  \bibfield  {author} {\bibinfo {author} {\bibfnamefont {J.}~\bibnamefont
  {Suzuki}},\ }\bibfield  {title} {\bibinfo {title} {Information geometrical
  characterization of quantum statistical models in quantum estimation
  theory},\ }\href {https://doi.org/10.3390/e21070703} {\bibfield  {journal}
  {\bibinfo  {journal} {Entropy}\ }\textbf {\bibinfo {volume} {21}},\ \bibinfo
  {pages} {703} (\bibinfo {year} {2019})}\BibitemShut {NoStop}%
\bibitem [{\citenamefont {Suzuki}\ \emph {et~al.}()\citenamefont {Suzuki},
  \citenamefont {Yang},\ and\ \citenamefont {Hayashi}}]{Suzuki2019a}%
  \BibitemOpen
  \bibfield  {author} {\bibinfo {author} {\bibfnamefont {J.}~\bibnamefont
  {Suzuki}}, \bibinfo {author} {\bibfnamefont {Y.}~\bibnamefont {Yang}},\ and\
  \bibinfo {author} {\bibfnamefont {M.}~\bibnamefont {Hayashi}},\ }\bibfield
  {title} {\bibinfo {title} {Quantum state estimation with nuisance
  parameters},\ }\href@noop {} {\ }\Eprint {https://arxiv.org/abs/1911.02790v3}
  {arXiv:1911.02790v3} \BibitemShut {NoStop}%
\bibitem [{\citenamefont {Kull}\ \emph {et~al.}(2020)\citenamefont {Kull},
  \citenamefont {Gu{\'{e}}rin},\ and\ \citenamefont {Verstraete}}]{Kull2020}%
  \BibitemOpen
  \bibfield  {author} {\bibinfo {author} {\bibfnamefont {I.}~\bibnamefont
  {Kull}}, \bibinfo {author} {\bibfnamefont {P.~A.}\ \bibnamefont
  {Gu{\'{e}}rin}},\ and\ \bibinfo {author} {\bibfnamefont {F.}~\bibnamefont
  {Verstraete}},\ }\bibfield  {title} {\bibinfo {title} {Uncertainty and
  trade-offs in quantum multiparameter estimation},\ }\href
  {https://doi.org/10.1088/1751-8121/ab7f67} {\bibfield  {journal} {\bibinfo
  {journal} {J. Phys. A: Math. Theor.}\ }\textbf {\bibinfo {volume} {53}},\
  \bibinfo {pages} {244001} (\bibinfo {year} {2020})}\BibitemShut {NoStop}%
\bibitem [{\citenamefont {Carollo}\ \emph {et~al.}(2020)\citenamefont
  {Carollo}, \citenamefont {Valenti},\ and\ \citenamefont
  {Spagnolo}}]{Carollo2020}%
  \BibitemOpen
  \bibfield  {author} {\bibinfo {author} {\bibfnamefont {A.}~\bibnamefont
  {Carollo}}, \bibinfo {author} {\bibfnamefont {D.}~\bibnamefont {Valenti}},\
  and\ \bibinfo {author} {\bibfnamefont {B.}~\bibnamefont {Spagnolo}},\
  }\bibfield  {title} {\bibinfo {title} {Geometry of quantum phase
  transitions},\ }\href
  {https://doi.org/https://doi.org/10.1016/j.physrep.2019.11.002} {\bibfield
  {journal} {\bibinfo  {journal} {Phys. Rep.}\ }\textbf {\bibinfo {volume}
  {838}},\ \bibinfo {pages} {1} (\bibinfo {year} {2020})}\BibitemShut {NoStop}%
\bibitem [{\citenamefont {Li}\ and\ \citenamefont {Luo}(2017)}]{Li2017}%
  \BibitemOpen
  \bibfield  {author} {\bibinfo {author} {\bibfnamefont {N.}~\bibnamefont
  {Li}}\ and\ \bibinfo {author} {\bibfnamefont {S.}~\bibnamefont {Luo}},\
  }\bibfield  {title} {\bibinfo {title} {Fisher concord: Efficiency of quantum
  measurement},\ }\href {https://doi.org/10.1515/qmetro-2016-0008} {\bibfield
  {journal} {\bibinfo  {journal} {Quantum Measurements and Quantum Metrology}\
  }\textbf {\bibinfo {volume} {3}},\ \bibinfo {pages} {44} (\bibinfo {year}
  {2017})}\BibitemShut {NoStop}%
\bibitem [{\citenamefont {Liu}\ \emph {et~al.}(2020)\citenamefont {Liu},
  \citenamefont {Yuan}, \citenamefont {Lu},\ and\ \citenamefont
  {Wang}}]{Liu2020}%
  \BibitemOpen
  \bibfield  {author} {\bibinfo {author} {\bibfnamefont {J.}~\bibnamefont
  {Liu}}, \bibinfo {author} {\bibfnamefont {H.}~\bibnamefont {Yuan}}, \bibinfo
  {author} {\bibfnamefont {X.-M.}\ \bibnamefont {Lu}},\ and\ \bibinfo {author}
  {\bibfnamefont {X.}~\bibnamefont {Wang}},\ }\bibfield  {title} {\bibinfo
  {title} {Quantum {Fisher} information matrix and multiparameter estimation},\
  }\href {https://doi.org/10.1088/1751-8121/ab5d4d} {\bibfield  {journal}
  {\bibinfo  {journal} {J. Phys. A: Math. Theor.}\ }\textbf {\bibinfo {volume}
  {53}},\ \bibinfo {pages} {023001} (\bibinfo {year} {2020})}\BibitemShut
  {NoStop}%
\bibitem [{\citenamefont {Gill}\ and\ \citenamefont {Massar}(2000)}]{Gill2000}%
  \BibitemOpen
  \bibfield  {author} {\bibinfo {author} {\bibfnamefont {R.~D.}\ \bibnamefont
  {Gill}}\ and\ \bibinfo {author} {\bibfnamefont {S.}~\bibnamefont {Massar}},\
  }\bibfield  {title} {\bibinfo {title} {State estimation for large
  ensembles},\ }\href {https://doi.org/10.1103/PhysRevA.61.042312} {\bibfield
  {journal} {\bibinfo  {journal} {Phys. Rev. A}\ }\textbf {\bibinfo {volume}
  {61}},\ \bibinfo {pages} {042312} (\bibinfo {year} {2000})}\BibitemShut
  {NoStop}%
\bibitem [{\citenamefont {Nagaoka}(2005)}]{Nagaoka2005a}%
  \BibitemOpen
  \bibfield  {author} {\bibinfo {author} {\bibfnamefont {H.}~\bibnamefont
  {Nagaoka}},\ }\bibinfo {title} {A new approach to {Cram\'er-Rao} bounds for
  quantum state estimation},\ in\ \href
  {https://doi.org/10.1142/9789812563071_0009} {\emph {\bibinfo {booktitle}
  {Asymptotic Theory of Quantum Statistical Inference}}},\ \bibinfo {editor}
  {edited by\ \bibinfo {editor} {\bibfnamefont {M.}~\bibnamefont {Hayashi}}}\
  (\bibinfo  {publisher} {World Scientific Publishing Co. Pte. Ltd.,
  Singapore},\ \bibinfo {year} {2005})\ pp.\ \bibinfo {pages}
  {100--112}\BibitemShut {NoStop}%
\bibitem [{\citenamefont {Matsumoto}(2002)}]{Matsumoto2002}%
  \BibitemOpen
  \bibfield  {author} {\bibinfo {author} {\bibfnamefont {K.}~\bibnamefont
  {Matsumoto}},\ }\bibfield  {title} {\bibinfo {title} {A new approach to the
  {Cram\'er-Rao}-type bound of the pure-state model},\ }\href
  {http://stacks.iop.org/0305-4470/35/i=13/a=307} {\bibfield  {journal}
  {\bibinfo  {journal} {J. Phys. A: Math. Gen.}\ }\textbf {\bibinfo {volume}
  {35}},\ \bibinfo {pages} {3111} (\bibinfo {year} {2002})}\BibitemShut
  {NoStop}%
\bibitem [{\citenamefont {Gill}\ and\ \citenamefont {Guţă}(2013)}]{Gill2013}%
  \BibitemOpen
  \bibfield  {author} {\bibinfo {author} {\bibfnamefont {R.~D.}\ \bibnamefont
  {Gill}}\ and\ \bibinfo {author} {\bibfnamefont {M.~I.}\ \bibnamefont
  {Guţă}},\ }\bibinfo {title} {On asymptotic quantum statistical inference},\
  in\ \href {https://doi.org/10.1214/12-IMSCOLL909} {\emph {\bibinfo
  {booktitle} {From Probability to Statistics and Back: High-Dimensional Models
  and Processes -- A Festschrift in Honor of Jon A. Wellner}}},\ \bibinfo
  {series} {IMS Collections}, Vol.~\bibinfo {volume} {9},\ \bibinfo {editor}
  {edited by\ \bibinfo {editor} {\bibfnamefont {M.}~\bibnamefont {Banerjee}},
  \bibinfo {editor} {\bibfnamefont {F.}~\bibnamefont {Bunea}}, \bibinfo
  {editor} {\bibfnamefont {J.}~\bibnamefont {Huang}}, \bibinfo {editor}
  {\bibfnamefont {V.}~\bibnamefont {Koltchinskii}},\ and\ \bibinfo {editor}
  {\bibfnamefont {M.~H.}\ \bibnamefont {Maathuis}}}\ (\bibinfo  {publisher}
  {Institute of Mathematical Statistics},\ \bibinfo {address} {Beachwood, Ohio,
  USA},\ \bibinfo {year} {2013})\ pp.\ \bibinfo {pages} {105--127}\BibitemShut
  {NoStop}%
\bibitem [{\citenamefont {Hayashi}\ and\ \citenamefont
  {Matsumoto}(2008)}]{Hayashi2008}%
  \BibitemOpen
  \bibfield  {author} {\bibinfo {author} {\bibfnamefont {M.}~\bibnamefont
  {Hayashi}}\ and\ \bibinfo {author} {\bibfnamefont {K.}~\bibnamefont
  {Matsumoto}},\ }\bibfield  {title} {\bibinfo {title} {Asymptotic performance
  of optimal state estimation in qubit system},\ }\href
  {https://doi.org/10.1063/1.2988130} {\bibfield  {journal} {\bibinfo
  {journal} {Journal of Mathematical Physics}\ }\textbf {\bibinfo {volume}
  {49}},\ \bibinfo {pages} {102101} (\bibinfo {year} {2008})},\ \Eprint
  {https://arxiv.org/abs/https://doi.org/10.1063/1.2988130}
  {https://doi.org/10.1063/1.2988130} \BibitemShut {NoStop}%
\bibitem [{\citenamefont {Kahn}\ and\ \citenamefont {Guţă}(2009)}]{Kahn2009}%
  \BibitemOpen
  \bibfield  {author} {\bibinfo {author} {\bibfnamefont {J.}~\bibnamefont
  {Kahn}}\ and\ \bibinfo {author} {\bibfnamefont {M.}~\bibnamefont {Guţă}},\
  }\bibfield  {title} {\bibinfo {title} {Local asymptotic normality for finite
  dimensional quantum systems},\ }\href
  {https://doi.org/10.1007/s00220-009-0787-3} {\bibfield  {journal} {\bibinfo
  {journal} {Communications in Mathematical Physics}\ }\textbf {\bibinfo
  {volume} {289}},\ \bibinfo {pages} {597} (\bibinfo {year}
  {2009})}\BibitemShut {NoStop}%
\bibitem [{\citenamefont {Yamagata}\ \emph {et~al.}(2013)\citenamefont
  {Yamagata}, \citenamefont {Fujiwara},\ and\ \citenamefont
  {Gill}}]{Yamagata2013}%
  \BibitemOpen
  \bibfield  {author} {\bibinfo {author} {\bibfnamefont {K.}~\bibnamefont
  {Yamagata}}, \bibinfo {author} {\bibfnamefont {A.}~\bibnamefont {Fujiwara}},\
  and\ \bibinfo {author} {\bibfnamefont {R.~D.}\ \bibnamefont {Gill}},\
  }\bibfield  {title} {\bibinfo {title} {Quantum local asymptotic normality
  based on a new quantum likelihood ratio},\ }\href
  {https://doi.org/10.1214/13-AOS1147} {\bibfield  {journal} {\bibinfo
  {journal} {Ann. Statist.}\ }\textbf {\bibinfo {volume} {41}},\ \bibinfo
  {pages} {2197} (\bibinfo {year} {2013})}\BibitemShut {NoStop}%
\bibitem [{\citenamefont {Gu\c{t}\u{a}}\ and\ \citenamefont
  {Kahn}(2006)}]{Guta2006}%
  \BibitemOpen
  \bibfield  {author} {\bibinfo {author} {\bibfnamefont {M.}~\bibnamefont
  {Gu\c{t}\u{a}}}\ and\ \bibinfo {author} {\bibfnamefont {J.}~\bibnamefont
  {Kahn}},\ }\bibfield  {title} {\bibinfo {title} {Local asymptotic normality
  for qubit states},\ }\href {https://doi.org/10.1103/PhysRevA.73.052108}
  {\bibfield  {journal} {\bibinfo  {journal} {Phys. Rev. A}\ }\textbf {\bibinfo
  {volume} {73}},\ \bibinfo {pages} {052108} (\bibinfo {year}
  {2006})}\BibitemShut {NoStop}%
\bibitem [{\citenamefont {Ozawa}(2003)}]{Ozawa2003}%
  \BibitemOpen
  \bibfield  {author} {\bibinfo {author} {\bibfnamefont {M.}~\bibnamefont
  {Ozawa}},\ }\bibfield  {title} {\bibinfo {title} {Universally valid
  reformulation of the {Heisenberg} uncertainty principle on noise and
  disturbance in measurement},\ }\href
  {https://doi.org/10.1103/PhysRevA.67.042105} {\bibfield  {journal} {\bibinfo
  {journal} {Phys. Rev. A}\ }\textbf {\bibinfo {volume} {67}},\ \bibinfo
  {pages} {042105} (\bibinfo {year} {2003})}\BibitemShut {NoStop}%
\bibitem [{\citenamefont {Ozawa}(2004)}]{Ozawa2004a}%
  \BibitemOpen
  \bibfield  {author} {\bibinfo {author} {\bibfnamefont {M.}~\bibnamefont
  {Ozawa}},\ }\bibfield  {title} {\bibinfo {title} {Uncertainty relations for
  joint measurements of noncommuting observables},\ }\href
  {https://doi.org/http://dx.doi.org/10.1016/j.physleta.2003.12.001} {\bibfield
   {journal} {\bibinfo  {journal} {Phys. Lett. A}\ }\textbf {\bibinfo {volume}
  {320}},\ \bibinfo {pages} {367 } (\bibinfo {year} {2004})}\BibitemShut
  {NoStop}%
\bibitem [{\citenamefont {Hall}(2004)}]{Hall2004}%
  \BibitemOpen
  \bibfield  {author} {\bibinfo {author} {\bibfnamefont {M.~J.~W.}\
  \bibnamefont {Hall}},\ }\bibfield  {title} {\bibinfo {title} {Prior
  information: How to circumvent the standard joint-measurement uncertainty
  relation},\ }\href {https://doi.org/10.1103/PhysRevA.69.052113} {\bibfield
  {journal} {\bibinfo  {journal} {Phys. Rev. A}\ }\textbf {\bibinfo {volume}
  {69}},\ \bibinfo {pages} {052113} (\bibinfo {year} {2004})}\BibitemShut
  {NoStop}%
\bibitem [{\citenamefont {Weston}\ \emph {et~al.}(2013)\citenamefont {Weston},
  \citenamefont {Hall}, \citenamefont {Palsson}, \citenamefont {Wiseman},\ and\
  \citenamefont {Pryde}}]{Weston2013}%
  \BibitemOpen
  \bibfield  {author} {\bibinfo {author} {\bibfnamefont {M.~M.}\ \bibnamefont
  {Weston}}, \bibinfo {author} {\bibfnamefont {M.~J.~W.}\ \bibnamefont {Hall}},
  \bibinfo {author} {\bibfnamefont {M.~S.}\ \bibnamefont {Palsson}}, \bibinfo
  {author} {\bibfnamefont {H.~M.}\ \bibnamefont {Wiseman}},\ and\ \bibinfo
  {author} {\bibfnamefont {G.~J.}\ \bibnamefont {Pryde}},\ }\bibfield  {title}
  {\bibinfo {title} {Experimental test of universal complementarity
  relations},\ }\href {https://doi.org/10.1103/PhysRevLett.110.220402}
  {\bibfield  {journal} {\bibinfo  {journal} {Phys. Rev. Lett.}\ }\textbf
  {\bibinfo {volume} {110}},\ \bibinfo {pages} {220402} (\bibinfo {year}
  {2013})}\BibitemShut {NoStop}%
\bibitem [{\citenamefont {Branciard}(2013)}]{Branciard2013}%
  \BibitemOpen
  \bibfield  {author} {\bibinfo {author} {\bibfnamefont {C.}~\bibnamefont
  {Branciard}},\ }\bibfield  {title} {\bibinfo {title} {Error-tradeoff and
  error-disturbance relations for incompatible quantum measurements},\ }\href
  {https://doi.org/10.1073/pnas.1219331110} {\bibfield  {journal} {\bibinfo
  {journal} {Proc. Natl. Acad. Sci. U.S.A.}\ }\textbf {\bibinfo {volume}
  {110}},\ \bibinfo {pages} {6742} (\bibinfo {year} {2013})}\BibitemShut
  {NoStop}%
\bibitem [{\citenamefont {Lu}\ \emph {et~al.}(2014)\citenamefont {Lu},
  \citenamefont {Yu}, \citenamefont {Fujikawa},\ and\ \citenamefont
  {Oh}}]{Lu2014}%
  \BibitemOpen
  \bibfield  {author} {\bibinfo {author} {\bibfnamefont {X.-M.}\ \bibnamefont
  {Lu}}, \bibinfo {author} {\bibfnamefont {S.}~\bibnamefont {Yu}}, \bibinfo
  {author} {\bibfnamefont {K.}~\bibnamefont {Fujikawa}},\ and\ \bibinfo
  {author} {\bibfnamefont {C.~H.}\ \bibnamefont {Oh}},\ }\bibfield  {title}
  {\bibinfo {title} {Improved error-tradeoff and error-disturbance relations in
  terms of measurement error components},\ }\href
  {https://doi.org/10.1103/PhysRevA.90.042113} {\bibfield  {journal} {\bibinfo
  {journal} {Phys. Rev. A}\ }\textbf {\bibinfo {volume} {90}},\ \bibinfo
  {pages} {042113} (\bibinfo {year} {2014})}\BibitemShut {NoStop}%
\bibitem [{\citenamefont {Hiai}\ and\ \citenamefont {Petz}(2014)}]{Hiai2014}%
  \BibitemOpen
  \bibfield  {author} {\bibinfo {author} {\bibfnamefont {F.}~\bibnamefont
  {Hiai}}\ and\ \bibinfo {author} {\bibfnamefont {D.}~\bibnamefont {Petz}},\
  }\href
  {http://gen.lib.rus.ec/book/index.php?md5=a8386c8bf1ffa93cea435a0a2c8ed31b}
  {\emph {\bibinfo {title} {Introduction to Matrix Analysis and
  Applications}}},\ \bibinfo {edition} {1st}\ ed.,\ Universitext\ (\bibinfo
  {publisher} {Springer International Publishing, Cham},\ \bibinfo {year}
  {2014})\BibitemShut {NoStop}%
\bibitem [{Sup()}]{SupplementalMaterial}%
  \BibitemOpen
  \href@noop {} {}\bibinfo {note} {See Supplemental Material for detailed
  derivations}\BibitemShut {NoStop}%
\bibitem [{\citenamefont {Wolf}()}]{WolfBook}%
  \BibitemOpen
  \bibfield  {author} {\bibinfo {author} {\bibfnamefont {M.~M.}\ \bibnamefont
  {Wolf}},\ }\bibfield  {title} {\bibinfo {title} {Quantum channels \&
  operations--guided tour},\ }\bibinfo {note} {online available at
  \url{http://www-m5.ma.tum.de/foswiki/pub/M5/Allgemeines/MichaelWolf/QChannelLecture.pdf}}\BibitemShut
  {NoStop}%
\bibitem [{\citenamefont {Lu}\ \emph {et~al.}(2015)\citenamefont {Lu},
  \citenamefont {Yu},\ and\ \citenamefont {Oh}}]{Lu2015}%
  \BibitemOpen
  \bibfield  {author} {\bibinfo {author} {\bibfnamefont {X.-M.}\ \bibnamefont
  {Lu}}, \bibinfo {author} {\bibfnamefont {S.}~\bibnamefont {Yu}},\ and\
  \bibinfo {author} {\bibfnamefont {C.~H.}\ \bibnamefont {Oh}},\ }\bibfield
  {title} {\bibinfo {title} {Robust quantum metrological schemes based on
  protection of quantum {Fisher} information},\ }\href
  {https://doi.org/10.1038/ncomms8282} {\bibfield  {journal} {\bibinfo
  {journal} {Nat. Commun.}\ }\textbf {\bibinfo {volume} {6}},\ \bibinfo {pages}
  {7282} (\bibinfo {year} {2015})}\BibitemShut {NoStop}%
\bibitem [{\citenamefont {Ozawa}(2019)}]{Ozawa2019}%
  \BibitemOpen
  \bibfield  {author} {\bibinfo {author} {\bibfnamefont {M.}~\bibnamefont
  {Ozawa}},\ }\bibfield  {title} {\bibinfo {title} {Soundness and completeness
  of quantum root-mean-square errors},\ }\href
  {https://doi.org/10.1038/s41534-018-0113-z} {\bibfield  {journal} {\bibinfo
  {journal} {npj Quantum Information}\ }\textbf {\bibinfo {volume} {5}},\
  \bibinfo {pages} {1} (\bibinfo {year} {2019})}\BibitemShut {NoStop}%
\bibitem [{\citenamefont {Ozawa}()}]{Ozaw2014}%
  \BibitemOpen
  \bibfield  {author} {\bibinfo {author} {\bibfnamefont {M.}~\bibnamefont
  {Ozawa}},\ }\bibfield  {title} {\bibinfo {title} {Error-disturbance relations
  in mixed states},\ }\href@noop {} {\ }\Eprint
  {https://arxiv.org/abs/1404.3388} {arXiv:1404.3388 [quant-th]} \BibitemShut
  {NoStop}%
\bibitem [{\citenamefont {Glauber}(1963)}]{Glauber1963a}%
  \BibitemOpen
  \bibfield  {author} {\bibinfo {author} {\bibfnamefont {R.~J.}\ \bibnamefont
  {Glauber}},\ }\bibfield  {title} {\bibinfo {title} {Coherent and incoherent
  states of the radiation field},\ }\href
  {https://doi.org/10.1103/PhysRev.131.2766} {\bibfield  {journal} {\bibinfo
  {journal} {Phys. Rev.}\ }\textbf {\bibinfo {volume} {131}},\ \bibinfo {pages}
  {2766} (\bibinfo {year} {1963})}\BibitemShut {NoStop}%
\bibitem [{\citenamefont {Busch}\ \emph {et~al.}(2014)\citenamefont {Busch},
  \citenamefont {Lahti},\ and\ \citenamefont {Werner}}]{Busch2014c}%
  \BibitemOpen
  \bibfield  {author} {\bibinfo {author} {\bibfnamefont {P.}~\bibnamefont
  {Busch}}, \bibinfo {author} {\bibfnamefont {P.}~\bibnamefont {Lahti}},\ and\
  \bibinfo {author} {\bibfnamefont {R.~F.}\ \bibnamefont {Werner}},\ }\bibfield
   {title} {\bibinfo {title} {Colloquium: Quantum root-mean-square error and
  measurement uncertainty relations},\ }\href
  {https://doi.org/10.1103/RevModPhys.86.1261} {\bibfield  {journal} {\bibinfo
  {journal} {Rev. Mod. Phys.}\ }\textbf {\bibinfo {volume} {86}},\ \bibinfo
  {pages} {1261} (\bibinfo {year} {2014})}\BibitemShut {NoStop}%
\bibitem [{\citenamefont {Appleby}(2016)}]{Appleby2016}%
  \BibitemOpen
  \bibfield  {author} {\bibinfo {author} {\bibfnamefont {D.~M.}\ \bibnamefont
  {Appleby}},\ }\bibfield  {title} {\bibinfo {title} {Quantum errors and
  disturbances: Response to {Busch}, {Lahti} and {Werner}},\ }\href
  {https://doi.org/10.3390/e18050174} {\bibfield  {journal} {\bibinfo
  {journal} {Entropy}\ }\textbf {\bibinfo {volume} {18}},\ \bibinfo {pages}
  {174} (\bibinfo {year} {2016})}\BibitemShut {NoStop}%
\bibitem [{\citenamefont {Provost}\ and\ \citenamefont
  {Vallee}(1980)}]{Provost1980}%
  \BibitemOpen
  \bibfield  {author} {\bibinfo {author} {\bibfnamefont {J.}~\bibnamefont
  {Provost}}\ and\ \bibinfo {author} {\bibfnamefont {G.}~\bibnamefont
  {Vallee}},\ }\bibfield  {title} {\bibinfo {title} {Riemannian structure on
  manifolds of quantum states},\ }\href {https://doi.org/10.1007/BF02193559}
  {\bibfield  {journal} {\bibinfo  {journal} {Commun. Math. Phys.}\ }\textbf
  {\bibinfo {volume} {76}},\ \bibinfo {pages} {289} (\bibinfo {year}
  {1980})}\BibitemShut {NoStop}%
\bibitem [{\citenamefont {Berry}(1989)}]{Berry1989}%
  \BibitemOpen
  \bibfield  {author} {\bibinfo {author} {\bibfnamefont {M.}~\bibnamefont
  {Berry}},\ }\bibinfo {title} {The quantum phase, five years after},\ in\
  \href@noop {} {\emph {\bibinfo {booktitle} {Geometric Phases in Physics}}},\
  \bibinfo {editor} {edited by\ \bibinfo {editor} {\bibfnamefont
  {A.}~\bibnamefont {Shapere}}\ and\ \bibinfo {editor} {\bibfnamefont
  {F.}~\bibnamefont {Wilczek}}}\ (\bibinfo  {publisher} {World Scientific,
  Singapore},\ \bibinfo {year} {1989})\ Chap.\ \bibinfo {chapter} {1.1}, pp.\
  \bibinfo {pages} {7--28}\BibitemShut {NoStop}%
\bibitem [{\citenamefont {Guo}\ \emph {et~al.}(2016)\citenamefont {Guo},
  \citenamefont {Zhong}, \citenamefont {Jing}, \citenamefont {Fu},\ and\
  \citenamefont {Wang}}]{Guo2016}%
  \BibitemOpen
  \bibfield  {author} {\bibinfo {author} {\bibfnamefont {W.}~\bibnamefont
  {Guo}}, \bibinfo {author} {\bibfnamefont {W.}~\bibnamefont {Zhong}}, \bibinfo
  {author} {\bibfnamefont {X.-X.}\ \bibnamefont {Jing}}, \bibinfo {author}
  {\bibfnamefont {L.-B.}\ \bibnamefont {Fu}},\ and\ \bibinfo {author}
  {\bibfnamefont {X.}~\bibnamefont {Wang}},\ }\bibfield  {title} {\bibinfo
  {title} {Berry curvature as a lower bound for multiparameter estimation},\
  }\href {https://doi.org/10.1103/PhysRevA.93.042115} {\bibfield  {journal}
  {\bibinfo  {journal} {Phys. Rev. A}\ }\textbf {\bibinfo {volume} {93}},\
  \bibinfo {pages} {042115} (\bibinfo {year} {2016})}\BibitemShut {NoStop}%
\bibitem [{\citenamefont {Xing}\ and\ \citenamefont {Fu}(2020)}]{Xing2020}%
  \BibitemOpen
  \bibfield  {author} {\bibinfo {author} {\bibfnamefont {H.}~\bibnamefont
  {Xing}}\ and\ \bibinfo {author} {\bibfnamefont {L.}~\bibnamefont {Fu}},\
  }\bibfield  {title} {\bibinfo {title} {Measure of the density of quantum
  states in information geometry and quantum multiparameter estimation},\
  }\href {https://doi.org/10.1103/PhysRevA.102.062613} {\bibfield  {journal}
  {\bibinfo  {journal} {Phys. Rev. A}\ }\textbf {\bibinfo {volume} {102}},\
  \bibinfo {pages} {062613} (\bibinfo {year} {2020})}\BibitemShut {NoStop}%
\bibitem [{\citenamefont {Arthurs}\ and\ \citenamefont
  {Kelly}(1965)}]{Arthurs1965}%
  \BibitemOpen
  \bibfield  {author} {\bibinfo {author} {\bibfnamefont {E.}~\bibnamefont
  {Arthurs}}\ and\ \bibinfo {author} {\bibfnamefont {J.~L.}\ \bibnamefont
  {Kelly}},\ }\bibfield  {title} {\enquote {\bibinfo {title} {On the
  simultaneous measurement of a pair of conjugate observables},}\ }\href@noop
  {} {\bibfield  {journal} {\bibinfo  {journal} {Bell Syst. Tech. J.}\ }\textbf
  {\bibinfo {volume} {44}},\ \bibinfo {pages} {725} (\bibinfo {year}
  {1965})}\BibitemShut {NoStop}%
\bibitem [{\citenamefont {Helstrom}(1973)}]{Helstrom1973a}%
  \BibitemOpen
  \bibfield  {author} {\bibinfo {author} {\bibfnamefont {Carl~W.}\ \bibnamefont
  {Helstrom}},\ }\bibfield  {title} {\enquote {\bibinfo {title}
  {{Cram\'er}-{Rao} inequalities for operator-valued measures in quantum
  mechanics},}\ }\href {\doibase 10.1007/BF00687093} {\bibfield  {journal}
  {\bibinfo  {journal} {Int. J. Theor. Phys.}\ }\textbf {\bibinfo {volume}
  {8}},\ \bibinfo {pages} {361--376} (\bibinfo {year} {1973})}\BibitemShut
  {NoStop}%
\bibitem [{\citenamefont {Garrison}\ and\ \citenamefont
  {Chiao}(2008)}]{Garrison2008}%
  \BibitemOpen
  \bibfield  {author} {\bibinfo {author} {\bibfnamefont {J.~C.}\ \bibnamefont
  {Garrison}}\ and\ \bibinfo {author} {\bibfnamefont {R.~Y.}\ \bibnamefont
  {Chiao}},\ }\href@noop {} {\emph {\bibinfo {title} {Quantum optics}}}\
  (\bibinfo  {publisher} {Oxford University Press, New York},\ \bibinfo {year}
  {2008})\BibitemShut {NoStop}%
\end{thebibliography}
\end{document}